\documentclass[10pt]{article}
\usepackage[OE]{express}
\usepackage{graphics}
\usepackage{dcolumn}
\usepackage{bm}
\usepackage[caption=false]{subfig}
\usepackage{xfrac}
\usepackage{mathrsfs}
\usepackage{float}
\usepackage[normalem]{ulem}
\usepackage[bbgreekl]{mathbbol}
\usepackage[utf8]{inputenc}
\usepackage{amsmath}
\usepackage{multirow}
\usepackage{adjustbox}
\usepackage{etoolbox}
\usepackage{comment}
\usepackage{amsmath}
\usepackage{amssymb}
\usepackage{blindtext}
\usepackage{hyperref}
\usepackage{cancel}

\renewcommand\thesubsection{\thesection.\Alph{subsection}}
\newcommand{\figref}[1]{Fig.~\ref{fig:#1}}
\newcommand{\figrefbegin}[1]{Figure~\ref{fig:#1}}

\newcommand{\secref}[1]{Sec.~\ref{#1}}
\newcommand{\secsref}[2]{Secs.~\ref{#1},\ref{#2}}

\renewcommand{\eqref}[1]{Eq.~(\ref{eq:#1})}
\newcommand{\eqrefbegin}[1]{Equation~(\ref{eq:#1})}
\newcommand{\eqsref}[2]{Eqs.~(\ref{eq:#1}) \mbox{and} (\ref{eq:#2})}
\newcommand{\eqssref}[2]{Eqs.~(\ref{eq:#1})--(\ref{eq:#2})}

\renewcommand{\vec}[1]{\mathbf{#1}}
\newcommand{\vecg}[1]{\boldsymbol{#1}}
\newcommand{\mat}[1]{\mathbb{#1}}
\newcommand{\Scale}[2][4]{\scalebox{#1}{$#2$}}

\newcommand{\subsecref}[1]{Sec.~\ref{subsec:#1}}
\newcommand{\appendixsec}[2]{\section{#1}\label{#2}}

\newcommand{\appendixref}[1]{appendix~\ref{#1}}
\newcommand{\unit}{1\!\!1}

\begin{document}
\title{General theory of spontaneous  emission near exceptional points}

\author{Adi Pick,\authormark{1,2,*} Bo Zhen,\authormark{1,3,4} Owen D. Miller,\authormark{5} Chia W. Hsu,\authormark{5} Felipe Hernandez,\authormark{6} Alejandro W. Rodriguez,\authormark{7} Marin Solja\v{c}i\'{c},\authormark{8} and Steven G. Johnson\authormark{6}}

\address{
\authormark{1}These authors contributed equally to this work.\\
\authormark{2}Department of Physics, Harvard University, Cambridge, Massachusetts 02138, USA\\
\authormark{3}Research Lab of Electronics, Massachusetts Institute of Technology, Cambridge, Massachusetts 02139, USA\\
\authormark{4}Physics Department and Solid State Institute, Technion, Haifa 32000, Israel\\
\authormark{5}Department of Applied Physics, Yale University, New Haven, Connecticut 06520, USA\\
\authormark{6}Department of Mathematics, Massachusetts Institute of Technology, 77 Massachusetts Avenue, Cambridge, Massachusetts 02139, USA\\
\authormark{7}Department of Electrical Engineering, Princeton University, Princeton, New Jersey 08544, USA\\
\authormark{8}Department of Physics, Massachusetts Institute of Technology, 77 Massachusetts Avenue, Cambridge, Massachusetts 02139, USA
}

\email{\authormark{*}adipick@physics.harvard.edu} 


\begin{abstract}
We present a general theory  of spontaneous  emission at exceptional points (EPs)---exotic degeneracies in non-Hermitian systems. Our theory extends beyond spontaneous emission to any light--matter interaction  described by the local density of states (e.g., absorption, thermal emission, and nonlinear frequency conversion).  
Whereas traditional spontaneous-emission  theories  imply infinite  enhancement factors  at EPs, we derive finite bounds on the enhancement, proving   maximum enhancement of 4 in passive systems with second-order EPs and  significantly larger enhancements (exceeding $400\times$) in gain-aided and higher-order EP systems.
In contrast to non-degenerate resonances, which are  typically associated with Lorentzian emission curves  in systems with low losses, EPs are associated with non-Lorentzian lineshapes, leading to  enhancements that scale nonlinearly  with the resonance quality factor. 
Our theory can be applied to dispersive media, with proper normalization of the resonant  modes.    
\end{abstract}




\section{Introduction}

Electromagnetic resonances enable precise control and enhancement of  spontaneous emission and other light--matter interactions. While it is well known that  resonances  can enhance spontaneous emission  rates via the celebrated  Purcell effect~\cite{Purcell1946,yokoyama1995,gaponenko2010} by  confining light to  small  volumes for long times,  recent work~\cite{firth2005giant,berry2003mode,lee2008divergent} suggests  that giant enhancements can occur via the less familiar Petermann effect~\cite{Petermann1979,van1996resonance,van1997excess,Siegman1989a,Siegman1989b}.  The  Petermann enhancement factor  is a measure of non-orthogonality of the modes in non-Hermitian systems and it appears to diverge when two  modes  coalesce at an   \emph{exceptional point} (EP)---an exotic  degeneracy in which two modes share  the same frequency and mode profile~\cite{Moiseyev2011,Heiss2012}.  In recent years, there has been an explosion of interest in EPs  due to the many interesting and counter-intuitive  phenomena associated with them, e.g.,  unidirectional reflection and transmission~\cite{lin2011unidirectional,peng2014parity,feng2013experimental},  topological mode switching~\cite{doppler2016dynamically,xu2016topological,heiss2016mathematical}, intrinsic single-mode lasing~\cite{feng2014single,hodaei2014parity}, and lasers with unconventional pump dependence~\cite{brandstetter2014reversing,liertzer2012pump,peng2014loss}.  An understanding of spontaneous emission at EPs is essential for their  implementation in  optical devices, but the existing  theory is limited to   one-dimensional~\cite{yoo2011quantum,heiss2014fano} or discrete oscillator systems~\cite{van2006eigenvector}.

In this paper, we present a general theory of spontaneous emission near EPs. Our theory extends beyond spontaneous emission to any light--matter interaction described by the local density of states  (LDOS) or, more precisely, any situation in which one analyzes the contribution of a given resonance to the emission of a source, such as narrowband thermal emission~\cite{de2012conversion,liu2011taming,laroche2006coherent,chinmay2015giant},   absorption~\cite{sondergaard2012plasmonic}, perfect coherent absorption~\cite{wan2011time,chong2010coherent,sun2014experimental},  and nonlinear harmonic generation~\cite{linnenbank2016second}.
Whereas traditional theories of spontaneous emission  imply infinite  enhancement factors  at EPs (since the Petermann factor diverges), we use a modified Jordan-form-based perturbation theory to derive (finite) bounds on the enhancement at second- and higher-order EP systems. We show that line narrowing leads to a maximum enhancement of 4 in passive systems with second-order EPs and  significantly larger enhancements (exceeding $400\times$) in gain-aided and higher-order EP systems. Our analytical results are presented in \secref{Green}, where we express the  emission rate at an EP   in  terms of the degenerate mode and its  corresponding  Jordan vector. 
This derivation assumes negligible dispersion, but we show in \appendixref{AppendixNew} that the effect of dispersion amounts to merely  modifying the normalization of the resonant modes, changing the results quantitatively but not qualitatively. 
Then, in \secref{Plasmonics}, we  demonstrate the implications of our theory via   a  concrete numerical example of coupled plasmonic resonators.   Motivated by the fact that  an  EP is associated with a double pole in the Green's function, we find  specific locations where the emission lineshape  becomes  a  squared Lorentzian, with peak  amplitude scaling as $Q^2$, where $Q$ is the resonance quality factor (a dimensionless measure of the resonance lifetime). We  show that the enhancement at the EP is thus, potentially, much larger than the Purcell factor, which scales linearly with   $Q$.   Then, in \secref{EnhancementLimits}, we derive     bounds on the maximal   enhancement  at an EP, and we explore  these bounds  using      a periodic  system,  which allows us to  tune gain, loss, and degeneracy independently.  
Our theory provides a quantitative prescription for achieving large enhancements in practical optical systems, which is applicable to arbitrary geometries and materials and can be implemented with  the recent experimental realizations of EPs~\cite{zhen2015,hodaei2014parity,feng2014single,feng2013experimental,gao2015observation,brandstetter2014reversing,doppler2016dynamically,xu2016topological}.

Traditional enhancement formulas fail at EPs since they are based on non-degenerate perturbation theory, which is invalid at EPs. Standard perturbation theory  relies on Taylor expansions of differentiable functions while, near EPs,  eigenvalues change non-analytically in response  to small matrix perturbations. Instead,  
one needs to use a Jordan-form-based perturbative expansion~\cite{Seyranian2003,brody2013information}. 
Although Jordan-vector perturbation theory is well known in linear algebra, along with related results on resolvent operators, these algebraic facts have not previously been applied to analyze Purcell/Petermann enhancement or LDOS in a general EP setting.
By using such a modified expansion, we obtain a quantitative  formula for the LDOS, which is  a measure of how much power a dipole source can radiate~\cite{Taflove2013} or, equivalently, a measure of its spontaneous emission rate.  Note that similar expansion  methods were previously used to evaluate the Green's functions at  EPs~\cite{Hanson2003,Hernandez2003,Hernandez2000}; however,  these works were limited to one-dimensional and paraxial systems, and were not applied to study   spontaneous emission.  
An alternative semi-analytic approach  was presented in \cite{yoo2011quantum}. In this work, the authors applied a scattering matrix formulation to 
model  a simple one-dimensional system   and analyzed its emission  properties  under $\mathcal{PT}$-symmetry conditions.  With proper modifications, such an analysis  could   be     generalized to handle  more complicated \emph{one-dimensional} structures (e.g.,  continuously varying  media or  complex layered media).
However, our general formulas [\eqsref{DoubleGreen}{DoubleLDOS}]  can be directly applied to \emph{any} system with an EP (e.g., three-dimensional photonic or plasmonic structures). Our results   demonstrate that the unique spectral properties at EPs are general  and do  not rely on certain symmetry or dimensionality.  Moreover, our  theory enables modeling complex experimental apparatuses, performing numerical optimization and design, and deriving bounds on the enhancement, thereby clarifying the usefulness and limitations of EPs for enhancing light matter interactions.

Formally, the Petermann  factor is inversely proportional to the ``unconjugated norm'' of the resonant  mode $\int\!dx\,\varepsilon\, \vec{E}_n^2$, where  $\vec{E}_n$ is the mode profile and $\varepsilon$ is the dielectric permittivity  (with modifications  to this ``norm'' when treating dispersive media~\cite{Jackson1999}.)
At an EP,  the unconjugated norm vanishes, $\int \!dx\,\varepsilon\, \vec{E}_n^2 = 0$~\cite{Heiss2012} (a property also called ``self-orthogonality''~\cite{Moiseyev2011}), and the Petermann factor  diverges.  In fact, the Petermann factor  can \emph{only} diverge at an EP.
This is because the Petermann factor is proportional to the sensitivity of an eigenvalue to perturbations~\cite{trefethen2005spectra}  (its ``condition number''), and that sensitivity can only diverge when two eigenvectors coalesce (i.e., at an EP) by the Bauer--Fike theorem~\cite{trefethen2005spectra}.  
This implies that our theory is applicable to any system exhibiting a giant Petermann factor. Specifically, in  \emph{any} laser and optical parametric oscillator (OPO)  system where a giant Petermann factor  was identified~\cite{longhi2000enhanced,d2009giantA,d2009giantB}, there must have been a nearby ``hidden'' EP.

\section{Local density of states and  Green's function expansions}
\label{Green}

In the following section, 
we give some background  on LDOS calculations in  non-degenerate systems, i.e., systems without EPs   (\subsecref{NonDegen}), and then we review perturbation theory for systems with   EPs  (\subsecref{DegenPert}). 
Finally, in \subsecref{LDOSDoublePole}, we present a condensed derivation of our key  analytical result---a   formula for the LDOS at an  EP [\eqsref{DoubleGreen}{DoubleLDOS}].

\subsection{LDOS formula for non-degenerate resonances}
\label{subsec:NonDegen}

The spontaneous emission rate of a dipolar emitter, oriented along the direction $\hat{\mathbf{e}}_\mu$, is proportional to the local density of states (LDOS)~\cite{wijnands1997green,xu2000quantum,sakoda2004optical}, 
which can be   related to   the   dyadic Green's function ${G}$  via~\cite{wijnands1997green,lagendijk1996resonant,Taflove2013}
\begin{align}
\mbox{LDOS}_\mu(\vec{x},\omega) = -\tfrac{2\omega}{\pi}\mbox{Im}[G_{\mu\mu}(\vec{x},\vec{x},\omega)].
\label{eq:LDOStoG}
\end{align} 
Here,   ${G}$ is defined as the response field to a  point source $\vec{J} = \delta(\vec{x}-\vec{x}')\hat{\mathbf{e}}_\mu$
at frequency $\omega$.
More generally,  currents and fields 
are related via Maxwell's frequency-domain partial differential equation,
$({\nabla\times\nabla\times-\,\omega^2\varepsilon})\vec{E} = i\omega\vec{J}$, where  $\varepsilon$ is the dielectric permittivity of the medium. 
Throughout the paper, we use bold letters for vectors, carets  for unit vectors, and  Greek letters for    vector components. Moreover, we set the speed of light to be one ($c=1$).

Computationally, one can directly invert Maxwell's equations to find ${G}$ and evaluate \eqref{LDOStoG}, but this provides little intuitive understanding. A modal expansion of the Green's function, when  applied properly, can be more insightful. Away from an EP, one can use the standard modal expansion formula for non-dispersive media~\cite{Arfken2006}:
\begin{align}
G_{\mu\mu}(\omega,\vec{x},\vec{x}') = 
\sum_n \frac{{E}_{n\mu}^R(\vec{x}){E}^L_{n\mu}(\vec{x}')}
{(\omega^2-\omega^2_{n})
(\bold{E}^L_{n},\bold{E}_{n}^R)}.
\label{eq:GreenExpand}
\end{align}
(We review the derivation of this formula for non-dispersive media in \appendixref{Appendix1} and treat dispersion effects in \appendixref{AppendixNew}).
Here, $\vec{E}_n^R$ is  a solution to the source-free Maxwell's equation with  outgoing boundary conditions or, more explicitly, is a   right eigenvector of the   eigenvalue problem:~$\hat{\mathcal{A}}\,\vec{E}^R_{n}=\omega_{n}^2 \vec{E}^R_{n}$~\cite{joannopoulos2011photonic} (with $\hat{\mathcal{A}}\equiv\varepsilon^{-1}\nabla\times\nabla\times$). Left modes ($\vec{E}_{n}^L$) are   eigenvectors of the transposed operator~$\hat{\mathcal{A}}^T\!\equiv\!\nabla\times\nabla\times\varepsilon^{-1}$.~
In reciprocal media $\varepsilon = \varepsilon^T$, and one can easily derive a simple relation between left and right eigenvectors: $\vec{E}_n^L = \varepsilon \vec{E}_n^R$.  
Right and left modes which correspond  to different eigenvalues  are orthogonal under the unconjugated ``inner product'' $(\bold{E}_{n}^L,\bold{E}_{m}^R)\equiv \int\!dx\,\vec{E}^L_{n} \cdot \vec{E}^R_{m} = \delta_{m,n}$~\cite{morse1953methods,siegman1986lasers,tureci2006self}.
[The convergence of the denominator $(\bold{E}^L_{n},\bold{E}_{n}^R)$ is proven in \appendixref{Appendix6}.]
Due to the outgoing boundary condition,  the modes solve a non-Hermitian eigenvalue problem  whose  eigenvalues ($\omega_n^2$) are generally   complex, with the imaginary part indicating the decay of modal energy in time (in accordance  with our intuition that typical resonances have finite  lifetimes). 
From \eqref{GreenExpand}, it follows   that  the  eigenfrequencies,  $\pm\omega_n$, are poles of the Green's function---a key concept in the mathematical analysis of resonances~\cite{newton2013scattering}. 
When considering  dispersive media, the denominator in  \eqref{GreenExpand} changes to 
$\int dx \vec{E}_n^L(x) [\omega^2\varepsilon(\omega,x) - \omega_n^2\varepsilon(\omega_n,x) ]\vec{E}_n^R(x)$, as shown in  
\appendixref{AppendixNew}.

In many cases of interest, one can get a fairly accurate approximation for the LDOS by including  only  low-loss resonances in the Green's function expansion [\eqref{GreenExpand}] since   only those   contribute substantially to the emission spectrum.  Under this approximation (i.e.,  considering only resonances $\omega_n=\Omega_n-i\gamma_n$ which lie close to the real axis in the complex plane, with $\gamma_n\ll\Omega_n$),  the spectral lineshape of the LDOS reduces to  a  sum of Lorentzian functions, weighted  by the local field intensity:
\begin{align}
\mbox{LDOS}_\mu(\vec{x},\omega) \approx
\sum_n
\frac{1}{\pi}
\frac{\gamma_n}{(\omega-\Omega_n)^2+\gamma_n^2}
\,\mbox{Re}\!\left[
\frac{{E}_{n\mu}^R(\vec{x}){E}^L_{n\mu}(\vec{x})}
{(\bold{E}^L_{n},\bold{E}_{n}^R)}\right]\!.
 \label{eq:LDOS_lorentzian}
\end{align}

Near the resonant frequencies, $\omega\approx\Omega_n$, the peak of the LDOS scales linearly with  the resonance quality factor $Q_n\equiv \frac{\Omega_n}{2\gamma_n}$, leading to the celebrated  Purcell enhancement factor~\cite{Purcell1946}.  
On the other hand, the ``unconjugated norm,'' $(\vec{E}_{n}^{L}, \vec{E}_{n}^{R})$,  which appears in  the denominator  of \eqref{LDOS_lorentzian}
 leads to the Petermann enhancement factor, defined as  ${({\vec{E}_{n}^R}^*\!,\vec{E}_{n}^R)({\vec{E}_{n}^L}^*\!,\vec{E}_{n}^L)}/ \left|(\vec{E}_{n}^L,\vec{E}_{n}^R)\right|^2$~\cite{berry2003mode}. 
In non-Hermitian systems, the mode profiles ($\vec{E}_n$) are complex and the Petermann factor is, generally, greater than one.
At the extreme case of an EP, the unconjugated norm in the denominator   vanishes and the  enhancement factor diverges. 
 However,   this divergence   does not properly describe LDOS or   spontaneous emission at EPs since
\eqref{LDOS_lorentzian} is  invalid at the EP.  That is  because the derivation of 
\eqref{GreenExpand} assumes that the set of eigenvectors of the operator $\hat{\mathcal{A}}$ spans  the Hilbert space, but this assumption breaks down at the  EP.  In order to complete the set of eigenvectors of $\hat{\mathcal{A}}$ into a basis and  obtain a valid  expansion for the Green's function and the LDOS  at the EP, we introduce in the following section additional  Jordan  vectors~\cite{trefethen2005spectra,Seyranian2003,weintraub2008jordan}.


\subsection{Jordan vectors and perturbation theory near EPs}
\label{subsec:DegenPert}

At a (second order) EP, the operator $\hat{\mathcal{A}}_0$ is defective---it does not have a complete basis of eigenvectors and is, therefore, not diagonalizable. However, one can find an eigenvector ($\vec{E}_0^R$) and an associated Jordan vector ($\vec{J}^R_0$), which satisfy the chain relations~\cite{Seyranian2003}:
 \begin{align}
&\hat{\mathcal{A}}_0{\vec{E}^R_0}=\lambda_{\mathrm{EP}}{\vec{E}^R_0},\nonumber \\
&\hat{\mathcal{A}}_0{\vec{J}^R_0}=\lambda_{\mathrm{EP}} {\vec{J}^R_0}+ {\vec{E}^R_0},
 \label{eq:Chain}
\end{align} 
where  $\lambda_{\mathrm{EP}} = \omega_\mathrm{EP}^2$ is the  degenerate eigenvalue. Equivalent expressions can be written   for the left eigenvector ${\vec{E}^L_0}$ and Jordan vector ${\vec{J}^L_0}$.  In order to uniquely define the Jordan chain vectors, we need to specify two normalization conditions, which we choose to be   $(\vec{E}^L_0,\vec{J}^R_0)=1$  and $(\vec{J}^L_0,\vec{J}^R_0) =0$.

Near the EP, on can find a  pair of nearly degenerate eigenvectors and eigenvalues that satisfy
\begin{align}
\hat{\mathcal{A}}(p){\vec{E}^R_\pm}=\lambda_\pm  {\vec{E}^R_\pm},
 \label{eq:SolvePT}
\end{align}
where $p\ll1$ represents a small deviation from the EP. 
[More explicitly, $\hat{\mathcal{A}}(p)\equiv \tfrac{1}{\varepsilon(p)}\nabla\times\nabla\times = \hat{\mathcal{A}}_0+\hat{\mathcal{A}}_1 p+\mathcal{O}(p^2)$, with  $\hat{\mathcal{A}}_0$ being  defective].
In order to obtain consistent perturbative expansions for $\vec{E}_\pm$ and $\lambda_\pm$ near the EP,  one can  use   alternating Puiseux series~\cite{Seyranian2003}:
 \begin{align}
&\lambda_\pm = \lambda_{0} \pm p^{\sfrac{1}{2}}\lambda_{1} + p\,\lambda_{2} \pm p^{\sfrac{3}{2}}\lambda_{3}+\hdots
 \nonumber\\
&\vec{E}_\pm^R = \vec{E}_{0}^R \pm p^{\sfrac{1}{2}} \,\vec{E}_{1}^R + p\,\vec{E}_{2}^R \pm p^{\sfrac{3}{2}}\,\vec{E}_{3}^R+\hdots
 \label{eq:ExpandVec}
\end{align}
Substituting \eqref{ExpandVec} into \eqref{SolvePT} and using the additional normalization condition $(\vec{J}_{0}^L,\vec{E}_\pm^R)=1$, one finds
that the  leading-order  terms in the series are 
\begin{align}
&\lambda_\pm = \lambda_{0} \pm p^{\sfrac{1}{2}}\Delta+\mathcal{O}(p),
\nonumber\\
&\vec{E}_\pm^R = \vec{E}_{0}^R \pm p^{\sfrac{1}{2}}\Delta\,\vec{J}_{0}^R+\mathcal{O}(p),
 \label{eq:AnsVec}
\end{align}
where $\Delta\!=\!\sqrt{\frac{(\vec{E}^L_0,\hat{\mathcal{A}}_1,\vec{E}^R_0)}{(\vec{J}^L_0,\vec{E}^R_0)}}$ and~
$(\vec{E}^L\!,\hat{\mathcal{A}}_1,\vec{E}^R)\!\equiv\!\int \vec{E}^L_0 \hat{\mathcal{A}}_1\vec{E}_0^R$. 
In the next section, we use these results to derive a formula for the LDOS at the EP.

 \subsection{LDOS  formula at  exceptional points}
 \label{subsec:LDOSDoublePole}

Near the EP (i.e., for small but non-zero $p$), one can use the non-degenerate  expansion formula \eqref{GreenExpand} to compute $G$. 
In order to compute  $G$ at the EP, we substitute   the perturbative expansions for $\lambda_\pm$ and $\vec{E}_\pm$   [\eqref{AnsVec}] into  \eqref{GreenExpand} and take    the limit of $p$ going  to zero, namely: 
 \begin{align}
 {G_{\mu\mu}^\mathrm{EP}(\omega,\vec{x},\vec{x}')\approx}
\displaystyle\lim_{p\rightarrow0}
&\left[
\frac{{E}_{+\mu}^R(\vec{x}){E}_{+\mu}^L(\vec{x'})}
{(\lambda-\lambda_+)(\vec{E}_+^L,\vec{E}_+^R)}
+\frac{{E}_{\mbox{-}\mu}^R(\vec{x}){E}_{\mbox{-}\mu}^L(\vec{x'})}
{(\lambda-\lambda_{\mbox{-}})(\vec{E}_{\mbox{-}}^L,\vec{E}_{\mbox{-}}^R)}
\right].
 \label{eq:Limit}
\end{align} 
The denominators in  \eqref{Limit} vanish in the limit of $p\rightarrow0$ since $(\vec{E}^L_\pm,\vec{E}^R_\pm)=\pm2\,p^{\frac{1}{2}}\lambda_1+\mathcal{O}(p)$ [as follows from \eqref{ExpandVec}]. 
Most importantly, however,    the   opposite signs   of the denominators   lead to cancellation of the  divergences and a finite  value remains, leading to 
\begin{align}
&G_{\mu\mu}^\mathrm{EP}(\omega,\vec{x},\vec{x}')\approx
\tfrac{1}{(\omega^2-\omega^2_\mathrm{\mathrm{EP}})^2}
\tfrac{{E}_{0\,\mu}^R(\vec{x}){E}_{0\,\mu}^L(\vec{x}')}
{(\vec{E}^L_0,\vec{J}^R_0)}+
\tfrac{1}{\omega^2-\omega^2_\mathrm{\mathrm{EP}}}
\tfrac{{E}_{0\,\mu}^R(\vec{x}){J}_{0\,\mu}^L(\vec{x}')+
{J}_{0\,\mu}^R(\vec{x}){E}_{0\,\mu}^L(\vec{x}')}
{(\vec{E}^L_0,\vec{J}^R_0)}.
\label{eq:DoubleGreen}
\end{align}
A formula for the LDOS  at the EP is obtained by taking the imaginary part of \eqref{DoubleGreen}.  Considering again  low-loss resonances (at which the enhancement is largest), and evaluating the  LDOS  near the EP frequency ($\omega\approx\omega_{\mathrm{EP}}$), we find
\begin{align}
&\mbox{LDOS}_{\mu}^\mathrm{EP}(\vec{x},\omega) \approx\nonumber\\
&\frac{\Omega_n}{2\pi}\left(\frac{\gamma_n}{(\omega-\Omega_n)^2+\gamma_n^2}\right)^2
\times \left[\frac{1}{2}
\mbox{Im}
\left(
\tfrac{{{E}_{0\,\mu}^R(\vec{x}){E}_{0\,\mu}^L(\vec{x})}}
{(\vec{E}_n^0,\vec{J}_n^0)}
\right)
-\frac{\omega-\Omega_n}{\gamma_n} 
\mbox{Re}
\left(
\tfrac{{{E}_{0\,\mu}^R(\vec{x}){E}_{0\,\mu}^L(\vec{x})}}
{(\vec{E}_n^0,\vec{J}_n^0)}
\right)
\right].
\label{eq:DoubleLDOS}
\end{align}

Crucially, \eqsref{DoubleGreen}{DoubleLDOS} yield finite results at the EP resonance frequency $\Omega_\mathrm{EP}$. Moreover, a squared Lorentzian prefactor, $\tfrac{\Omega_n}{2\pi}(\tfrac{\gamma_n}{(\omega-\Omega_n)^2+\gamma_n^2})^2$,  replaces the traditional Lorentzian prefactors near non-degenerate resonances, $\tfrac{1}{\pi}\tfrac{\gamma_n}{(\omega-\Omega_n)^2+\gamma_n^2}$ [compare with \eqref{LDOS_lorentzian}]. This unique spectral feature follows directly from the existence of a double pole in the modified expansion formula for $G$ [the first term in \eqref{DoubleGreen}].  As shown in the following section, a squared Lorentzian lineshape implies a narrower emission peak and greater resonant enhancement in comparison with a non-degenerate resonance at the same complex frequency.   The strength of our  formulation [\eqsref{DoubleGreen}{DoubleLDOS}] is that it is applicable to arbitrary structures and can therefore be used to design and optimize complex three-dimensional geometries with EPs. Moreover, it clarifies the conditions under which LDOS enhancement is maximal; essentially, determined by  the relative magnitude of the two terms in  \eqref{DoubleGreen}, depending  on  the location of the emitter.


\section{Properties of the LDOS at   EPs}
 \label{Plasmonics}
 
In this section, we explore the spectral emission properties at EPs [following \eqsref{DoubleGreen}{DoubleLDOS}] via a numerical model system of coupled plasmonic resonators.

\subsection{Example: EPs in plasmonic  systems}

\begin{figure}[h]
                 \centering\includegraphics[width=0.8\textwidth]{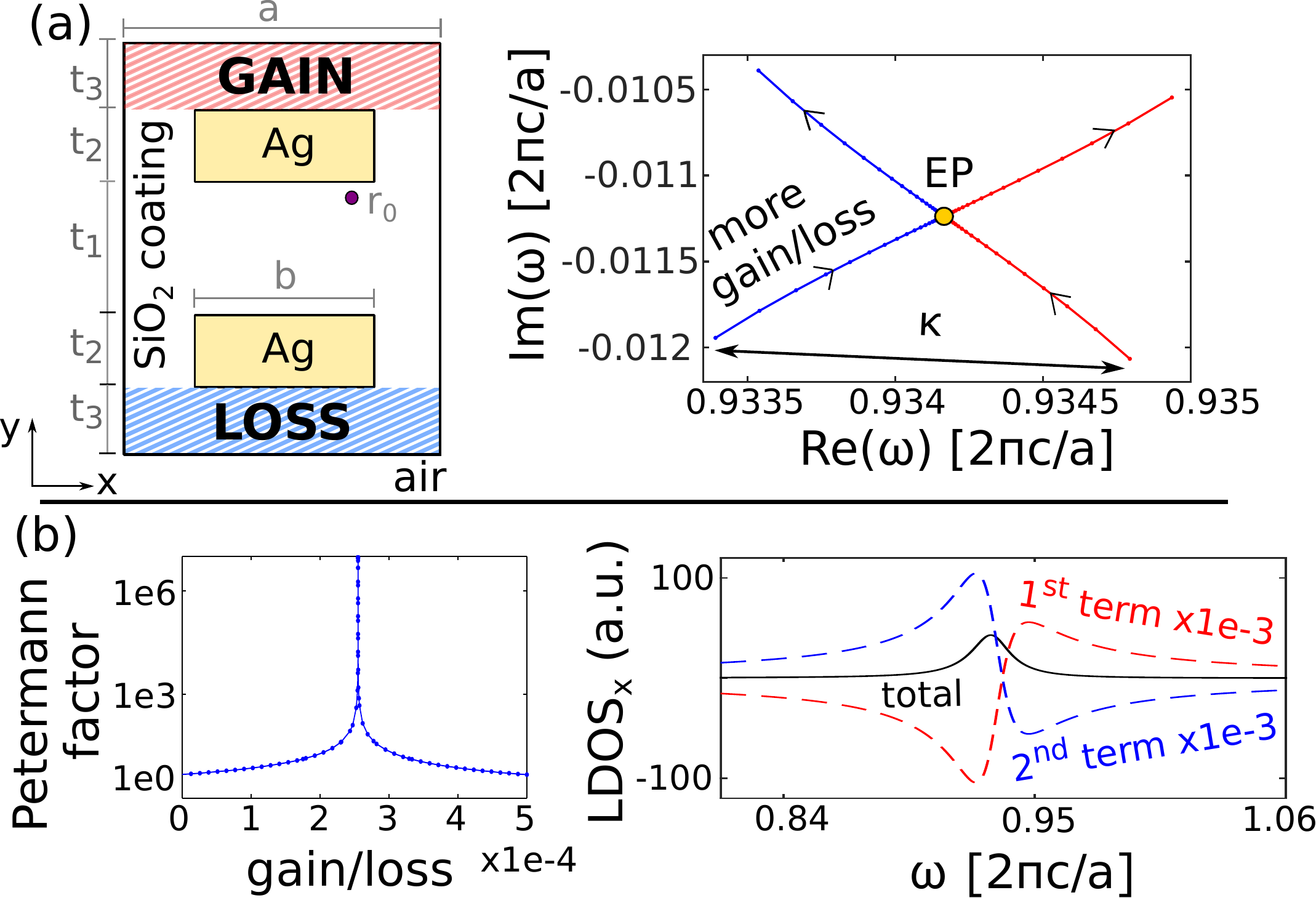}
                  \caption{Plasmonic system  with  exceptional points. 
(a) Left: Two   plasmonic resonators with  silver cores and a silica coating, with   gain and loss (red/blue) added to  the outer sides of the coating. Right: Eigenvalue trajectories in the complex plane upon increasing gain/loss. The trajectories merge at the EP  (orange dot). 
(b) The Petermann factor, ${({\vec{E}_{n}^R}^*\!,\vec{E}_{n}^R)({\vec{E}_{n}^L}^*\!,\vec{E}_{n}^L)}/ \left|(\vec{E}_{n}^L,\vec{E}_{n}^R)\right|^2$, diverges at the EP (left) while the LDOS$_x$ [evaluated  using \eqref{GreenExpand}] remains finite (right), since the giant contributions of the terms in \eqref{GreenExpand} have opposite signs (blue/red curves, scaled by $10^{-3}$).}
                  \label{fig:Structure}
  \end{figure}

A convenient approach for tailoring   the LDOS in practice is  by using  plasmonic resonances  in metallic structures, which can enable   ultra-high LDOS enhancements and are widely used in various applications~\cite{maier2007plasmonics}. In this subsection, we numerically   explore    a system  of two  plasmonic resonators with approximate  parity-time ($\mathcal{PT}$) symmetry.  $\mathcal{PT}$ symmetric  systems  in optics are characterized by having   balanced distributions of gain and loss, and are known to possess  EPs  at critical gain/loss values at which the mode profiles  undergo spontaneous symmetry breaking~\cite{ruter2010observation}. Our numerical setup is shown on the left-hand side of  \figref{Structure}(a). It includes  two rods with metallic cores ($\varepsilon=-2.3+0.0001i$) and a silica coating ($\varepsilon=2.1316$) surrounded by air ($\varepsilon=1$).
The dimensions of the structure ($b = 0.643a, t_1 = 0.536a,  t_2 =  0.16a$, and $t_3 = 0.268a$) were tuned in order to  have two nearly degenerate resonances that are spectrally separated from the neighboring resonances of the structure.
We implement outgoing boundary conditions    using perfectly matched layers (PML)~\cite{Taflove2013}.
In order to preserve the approximate $\mathcal{PT}$ symmetry of the system,  gain and loss are added symmetrically  to the outer parts of the coating. By brute-force optimization, we find that an EP occurs when the gain/loss value is $\left|\mathrm{Im}\,\varepsilon\right|\approx0.0002551$ and  the background permittivity of the upper half of the silica coating is slightly shifted to $\varepsilon \approx 2.129$.
The right-hand side of \figref{Structure}(a) depicts the trajectories of the two eigenvalues (red and blue curves) that merge at the EP (orange dot) upon varying the gain/loss parameter $\left|\mathrm{Im}\,\varepsilon\right|$. More details on the numerical optimization procedure are given in \appendixref{Appendix3}.

We discretize Maxwell's equations using a finite-difference frequency-domain (FDFD) approach~\cite{christ1987three} and  
 compute the LDOS by taking imaginary part of the Green's function. The Green's function is found via three methods: 
(i) directly inverting Maxwell's partial differential equation, 
(ii) using the non-degenerate expansion formula, \eqref{GreenExpand}, which is valid  away from the EP, and
(iii) using the  degenerate formula, \eqref{DoubleGreen},  at the EP.
In principle, one can  use the non-degenerate formula, \eqref{GreenExpand}, very close to the EP to compute the LDOS, relying on cancellation between the terms to yield a finite result. However, non-degenerate perturbation theory obfuscates the finite enhancement at the EP, confusing previous estimates of LDOS, whereas our formulation,  based on degenerate perturbation theory, naturally captures the finite enhancement.

Our numerical results are shown in  \figref{Structure}(b). 
The plot on the left shows the Petermann factor, which diverges at the EP, while the right-hand side plot shows the finite  LDOS. We excite TM modes and, therefore, compute the LDOS for x-polarized modes (denoted by LDOS$_x$). The red and blue curves show the two terms that contribute to the sum in \eqref{GreenExpand}. Upon approaching the EP, the individual  contributions to the sum diverge with opposite signs, while  their sum  (black curve) remains finite. The red and blue curves are scaled by $10^{-3}$ for ease of presentation.

\subsection{Simplified model for the LDOS}
 \label{subsec:simplifiedLDOS}

Although our general  formula for  the LDOS at the EP  can be directly applied to the  plasmonic system of \figref{Structure}, it is useful to introduce a simplified model to interpret the results.  
First, we   project  Maxwell's operator, $\hat{\mathcal{A}} = \varepsilon^{-1}\nabla\times\nabla\times$, onto  the two-dimensional subspace spanned by the nearly degenerate eigenvectors near the EP, thus producing a reduced   $2\times2$  matrix,  $\hat{A}$ (which gives a meaningful description of the system as long as the emission spectrum is dominated by the two coalescing  resonances).  
Second,  we employ an approach  similar in spirit to coupled-mode theory (CMT)~\cite{huang1994coupled,okamoto2010fundamentals}, which 
involves expressing  the modes of the  two-rod system (\emph{the coupled system}) in terms of modes  of two reduced systems (\emph{the uncoupled systems}), containing only one or the other rod. 
Such an approach is valid  whenever   the  rod--rod separation ($t_1$) in the coupled system is large enough so that the frequency  splitting induced by the coupling [$\kappa$ depicted in \figref{Structure}(a)] is smaller than the uncoupled   resonance frequencies.  
Denoting by $U$ and $V$ the matrices whose columns are  the right and left eigenvectors of the   uncoupled system ($\vec{E}^{R}_{1,2}$ and $\vec{E}^{L}_{1,2}$ corresponding to rod $1$ and $2$ respectively),  we find in \appendixref{Appendix2} that the reduced operator is 
\begin{align}
\hat{A} = V^T\hat{\mathcal{A}}\,U =
\left( \begin{array}{cc}
[\omega_\mathrm{EP}-i\eta]^2 & 2\Omega_\mathrm{EP}\kappa \\
2\Omega_\mathrm{EP}\kappa & [\omega_\mathrm{EP}+i\eta]^2
 \end{array} \right).
\label{eq:TCMTrestriction}
\end{align}
Here,  $\omega_\mathrm{EP}\equiv \Omega_\mathrm{EP}-i\gamma$ is the degenerate  eigenvalue  ($\Omega_\mathrm{EP}$ is the resonant  frequency while  $\gamma$ denotes the decay rate), and  $\kappa$ is the  near-field coupling between the rods. In general, $\kappa$ is complex; however, in  low-loss systems, $\kappa$ is almost real and, in the current analysis, we neglect its imaginary part entirely for ease of discussion. 
Finally, $\eta\approx\tfrac{\Omega_{\mathrm{EP}}}{2}\frac{\int\!dx\, \vec{E}_n^L(\mathrm{Im}\,\varepsilon/\varepsilon)\vec{E}_{n}^R}{(\vec{E}_n^L,\vec{E}_{n}^R)}$ is the imaginary frequency shift induced by  the gain and loss ($\mathrm{Im}\,\varepsilon$) in the coating.
(This definition of  $\eta$   follows from perturbation theory for small gain/loss~\cite{joannopoulos2011photonic}.) 
This approach is closely related to the recent $2\times2$ formalism used in $\mathcal{PT}$-symmetry works~\cite{peng2014parity,feng2013experimental,brandstetter2014reversing,hodaei2014parity,feng2014single}. In fact, the formulations are equivalent for low-loss resonances, $\gamma\ll\Omega_\mathrm{EP}$, which is the regime  considered in this section.    We note that the EP occurs at a complex frequency (i.e., below the real axis in the complex plane) due to outgoing boundary conditions.

Next, we obtain a simplified formula for the LDOS at the EP. 
The reduced  matrix $\hat{A}$ has an   EP at the     critical gain/loss value: $\eta = \kappa$. Denoting the   defective matrix  by $\hat{A}_0$ and the identity opertor by $\unit$,   we can  relate 
the full Green's function at the EP  to  $\hat{A}_0$ via:
\begin{align}
G_\mathrm{EP}(\vec{r},\vec{r}',\omega)\equiv
(\hat{\mathcal{A}}_0-\omega^2\Scale[1.2]{\unit})^{-1}\approx 
U(\hat{A}_0-\omega^2\Scale[1.2]{\unit})^{-1}V^T.
\label{eq:GtwoBYtwo}
\end{align}
A formula for the  LDOS  is obtained by taking the imaginary part of \eqref{GtwoBYtwo}, which allows expressing the LDOS in terms of entries of the $2\times2$ resolvent operator, 
$(\hat{A}_0-\omega^2\Scale[1.2]{\unit})^{-1}$, weighted by a product of the left and right local fields $\vec{E}_n^R\vec{E}_m^L$ (with $n,m = 1,2$ denoting the two resonances of the uncoupled system). 

\begin{figure}[h]  
	         \centering\includegraphics[width=0.7\textwidth]{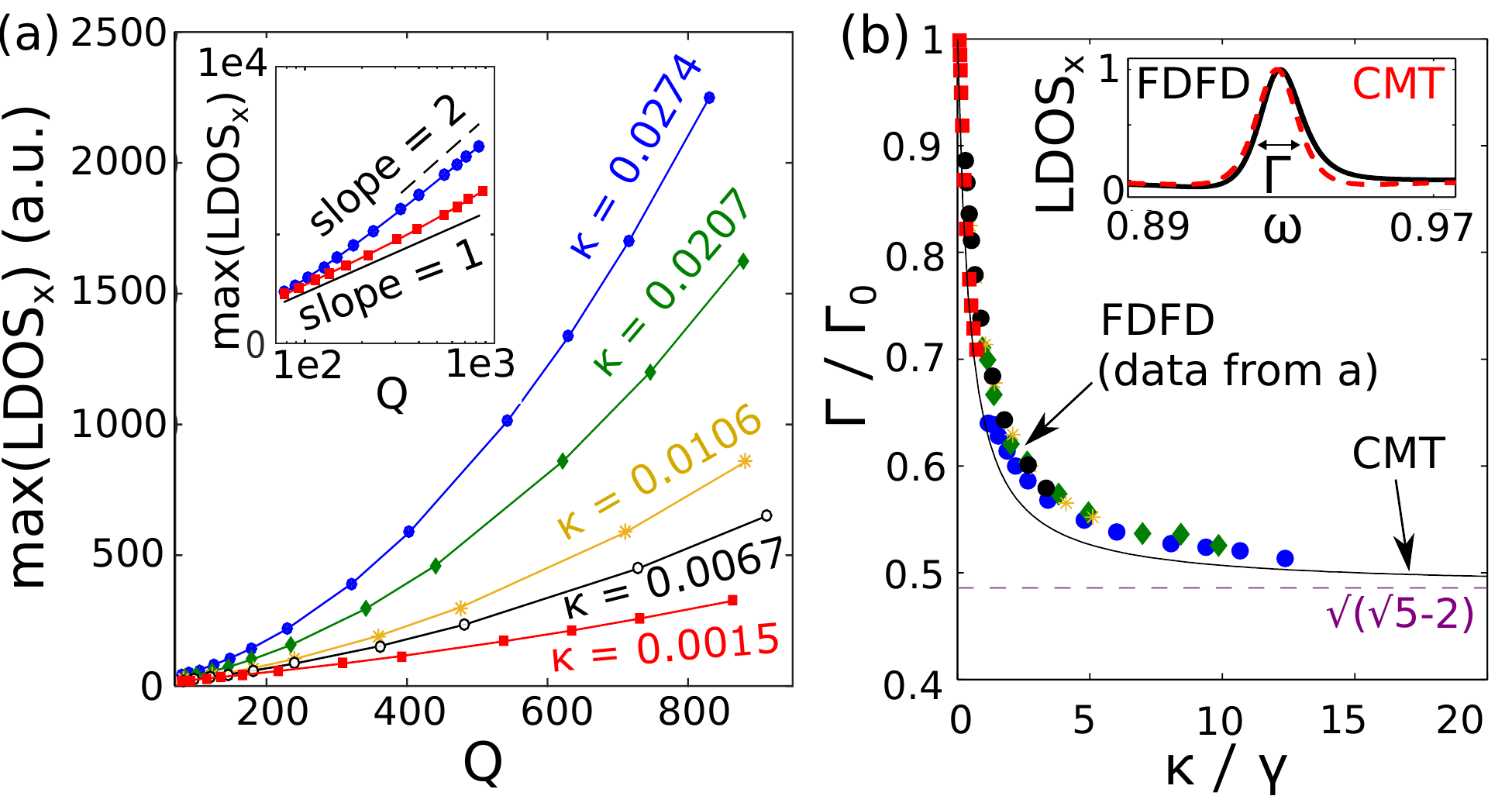}
                  \caption{
Spectral properties of  LDOS at EPs.  (a)  LDOS peak vs.  quality factor $Q$ for five  coupling values $\kappa$,   showing   quadratic (linear) scaling for  large (small) $\kappa$ values. 
$Q$ is varied by adding   gain to the coating, while  $\kappa$ is varied by changing  the rod--rod separation. Inset: Red and blue  points from the main plot, on a log-log scale.  	
(b) 
Normalized linewidth (FWHM)   at the EP, $\Gamma/\Gamma_0$, vs. normalized coupling, $\kappa/\gamma$ , computed using: 1.   FDFD-discretization of  Maxwell's equations  (dots), and  2. the  CMT-based linewidth formula, \eqref{FWHM_tcmt} (black line). 
The   limit as $\kappa/\gamma\rightarrow\infty$ is shown in purple (dashed line). 
($\Gamma_0\equiv2\gamma$ is the FWHM of a Lorentzian curve, see text.)
Inset:  LDOS$_x$ at the EP, computed via  FDFD (black) and via  CMT [\eqref{LDOSexplicit}, red dashed line].  
}
\label{fig:Qsqr}
  \end{figure}

The advantage of this formulation becomes apparent when evaluating the  LDOS   in close proximity to one of the  resonators.  For instance,   near the gain region [e.g., at  $\vec{r}_0$ in \figref{Structure}(a)], 
the lossy mode intensity, $\propto|\vec{E}_2^R\vec{E}_2^L|$, is  negligible,  so it  follows from \eqref{GtwoBYtwo} that  the LDOS is  proportional to the first diagonal entry of the resolvent
\begin{align} 
(\hat{A}_0-\omega^2\Scale[1.2]{\unit})^{-1}_{[1,1]}= 
\frac{1}{\omega^2-\omega^2_\mathrm{EP}}+
\frac{2\Omega_\mathrm{EP}\kappa}{[\omega^2-\omega^2_\mathrm{EP}]^2}.
\label{eq:resolvent}
\end{align}
Moreover, since we consider low-loss resonances,
we can normalize the mode profiles so that they are mainly real 
 ($\mathrm{Re}[\vec{E}_{1}]\approx \vec{E}_{1}$), and  we  find 
\begin{align}
&\mathrm{LDOS}_x(\vec{r}_0,\omega)\approx 
\tfrac{\vec{E}^R_1(\vec{r}_0)\vec{E}^L_1(\vec{r}_0)}{2\Omega_\mathrm{EP}}
\left[\tfrac{\gamma}{(\omega-\Omega_\mathrm{EP})^2+\gamma^2}
+\tfrac{\kappa\left[\gamma^2 - (\omega-\Omega_\mathrm{EP})^2\right]}
{\left[\gamma^2 - (\omega-\Omega_\mathrm{EP})^2\right]^2+\left[2\gamma(\omega-\Omega_\mathrm{EP})\right]^2}\right].
\label{eq:LDOSexplicit}
\end{align}
The inset in  \figref{Qsqr}(b) demonstrates the nearly perfect  agreement between this  simplified CMT-based LDOS formula  (red dashed curve) and  brute-force inversion of Maxwell's equation, discretized via   FDFD  (black curve).

\subsection{Quadratic scaling and linewidth narrowing}

In this subsection, we apply  our CMT-based simplified formulas [\eqssref{GtwoBYtwo}{LDOSexplicit}] to
evaluate the LDOS$_x$  near the upper rod [see \figref{Structure}(a)].
As can be  seen from \eqref{LDOSexplicit}, the   LDOS peak value  scales as 
\begin{align}
M_\mathrm{EP} \equiv
\max_\omega \{\mathrm{LDOS}(\vec{r}_0,\omega)\}
\propto \frac{1}{\gamma} + \frac{\kappa}{\gamma^2}.
\label{eq:Qsqr_eq}
\end{align}
When the resonance-decay rate $\gamma$ is much smaller than the mode-coupling rate $\kappa$, the lineshape   approaches a squared Lorentzian curve  and   $M_\mathrm{EP}$  scales quadratically with the quality factor $Q\equiv\tfrac{\Omega_\mathrm{EP}}{2\gamma}$~\cite{joannopoulos2011photonic}. On the other hand, when $\gamma\gg\kappa$, the LDOS peak  scales linearly with $Q$ (similar to  isolated resonances, as predicted by Purcell~\cite{Purcell1946}).
\figrefbegin{Qsqr}(a) demonstrates this key feature. To this end, we computed the  LDOS peak, $M_\mathrm{EP}$, for several  
resonance decay rates $\gamma$ (corresponding to several  $Q$ values), varied   by introducing homogeneous background gain   in the coating.   
We repeated this procedure for five rod--rod separations (corresponding to five $\kappa$ values,  ranging from $0.0015$ to $0.0274$), and verified the scaling laws of \eqref{Qsqr_eq}.
The inset in \figref{Qsqr}(a)  presents the blue and red data points  from  the main plot on a log-log scale, providing additional  confirmation  for  the quadratic scaling at $\gamma\ll \kappa$ (blue) and the linear scaling in the opposite limit of $\gamma\gg\kappa$ (red).

Another consequence of a squared Lorentzian lineshape is a   narrower emission peak, compared to that of a standard Lorentzian  spectrum. 
The   full-width half maximum (FWHM) of a  standard Lorentzian curve, $\tfrac{1}{\pi}\tfrac{\gamma_n}{(\omega-\Omega_n)^2+\gamma_n^2}$, is 
 $\Gamma_0 = 2\gamma$ (where  $\Omega_0 \pm\Gamma_0/2$ is the frequency at which the Lorentzian drops to half of its maximal value). 
 On the other hand, following \eqref{LDOSexplicit}, the FWHM of the LDOS near an EP with  $\mathrm{Im}[\omega_\mathrm{EP}] = \gamma$  is
\begin{align}
\Gamma = 
\Gamma_0\,
\sqrt{\frac{\sqrt{\gamma^2+2\gamma\kappa+5\kappa^2}-2\kappa}
{\gamma+\kappa}}.
\label{eq:FWHM_tcmt}
\end{align}
As shown  in   \figref{Qsqr}(b), the FWHM  at the EP  is always smaller than $\Gamma_0$, approaching a value of $\sqrt{\sqrt{5}-2}\Gamma_0\approx\,0.48\Gamma_0$ in the  limit of $\kappa/\gamma\rightarrow\infty$. We computed the FWHM 
directly from the FDFD-discretization of  Maxwell's equations (dots) and by using the CMT-based simplified expression, \eqref{LDOSexplicit} (black solid  line), proving very good  agreement between the two methods.


 \section{LDOS enhancement at EPs}
 \label{EnhancementLimits}
 
 In the previous section, we showed that a squared Lorentzian lineshape can lead to enhanced emission  rates and reduced linewidth. In this section, we quantify the enhancement at the   EP and study its bounds. 
To demonstrate our results, we employ another numerical example: a periodic waveguide (\subsecref{PeriodicStructure}). This structure  allows us to independently tune gain/loss and degeneracy and, therefore,   demonstrate the impacts of the two effects separately. 
We treat both   passive (\subsecref{PassiveEP2})  and active (\subsecref{ActiveEP2}) systems (i.e., systems without  and with gain respectively), and then generalize our results (in \subsecref{PassiveEP3}) to   higher-order  EPs, which form at the coalescence of  multiple  resonance.  

 \subsection{Example: EPs in periodic structures}
 \label{subsec:PeriodicStructure}

\begin{figure}[t]
                 \centering\includegraphics[width=1.0\textwidth]{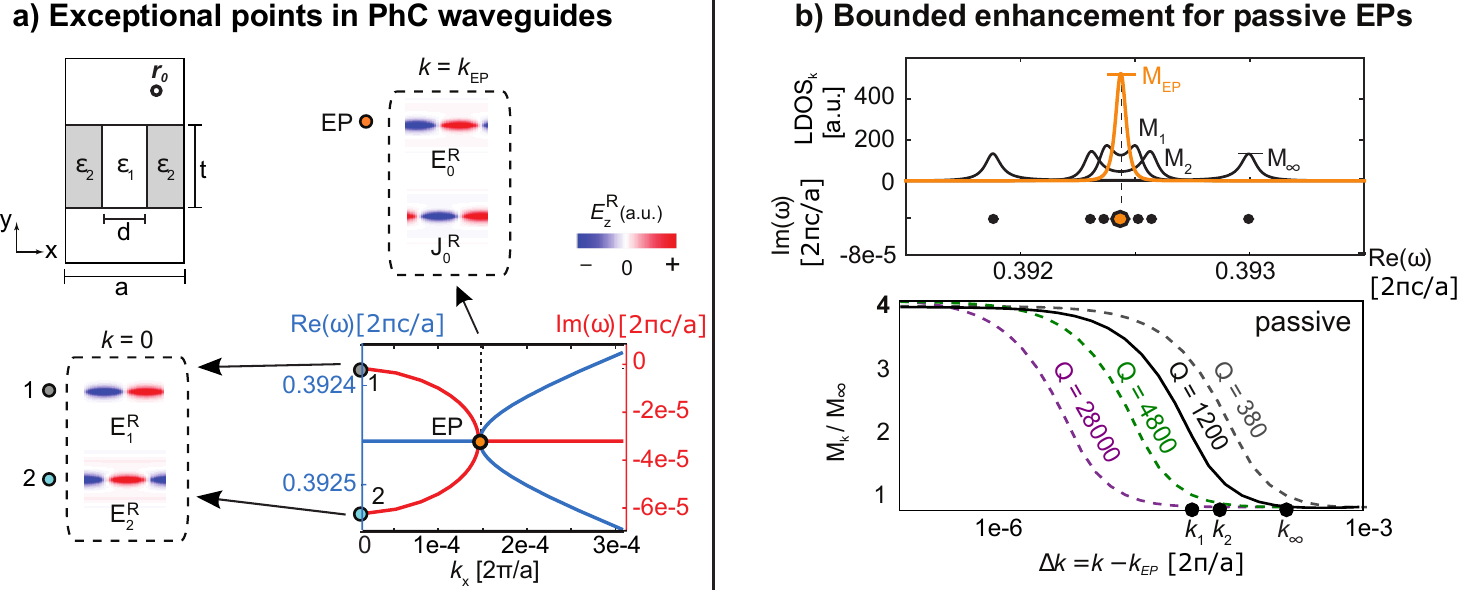}
                                   \caption{Passive periodic waveguides with EPs.
(a)  Top left: Periodic waveguide with outgoing boundary condition in the  transverse direction.  
Bottom left: Field patterns  of TM modes  at  $k=0$~(only the real part is shown).
Top right:  Degenerate mode  and Jordan vector at  $k_\mathrm{EP}$.
Bottom right: Resonance [$\omega_n(k)$] vs.  $k$-vector, showing an EP (orange dot) at   $k_\mathrm{EP}a/2\pi\approx1.46\!\times\!10^{-4}$. 
(b) Top plot, positive $y$-axis: 
LDOS$_k$ at $\vec{r}_0$ for four $k$-values ($k_\mathrm{EP}$, $k_{1}$, $k_{2}$, $k_\infty$,  marked on the lower plot).  
Negative $y$-axis: Resonances  [$\omega_n(k)$] in the complex plane. Bottom: Normalized LDOS$_k$ peak ($M_k/M_\infty$)  vs. deviation from the EP ($\Delta k$) for  four structures with different  $Q$ values, showing 4-fold enhancement at the EP. ($M_k$ is the LDOS peak at $k$.)
}
\label{fig:slab-passive}
  \end{figure}

To  demonstrate our results in this  section, we numerically explore   the periodic structure   shown in \figref{slab-passive}(a).  
The modes of a periodic system are Bloch wavefunctions  of the form~$\vec{E}(r) =\vec{E}_k(r)e^{i\vec{k}\cdot\vec{r}}$, where~$\vec{E}_k(r)$~is a periodic function and~$\vec{k}$~is the  wavevector~\cite{ashcroft1976solid}. 
At each $\vec{k}$, the mode $\vec{E}_k(r)$ solves an  eigenvalue problem of the form~\cite{huang1994coupled,okamoto2010fundamentals}:~$\hat{\mathcal{A}}(\vec{k})\vec{E}_{nk}=\omega_{nk}^2\vec{E}_{nk}$, where $\hat{\mathcal{A}}(\vec{k})\equiv \varepsilon^{-1}(\nabla + i\vec{k})\times(\nabla + i\vec{k})\times$ and $n$ labels the band. 
The figure of merit for spontaneous emission in  periodic structures is the  LDOS  per  wavevector $k$ and field component $\mu$, which is a measure of the  power expended by a Bloch-periodic dipole source with a particular $k$-vector, in the presence of an electromagnetic field polarized along direction $\mu$. We abbreviate it as LDOS$_k$ (also called the mutual density of states~\cite{mcphedran2004density}).  The LDOS$_k$ can be integrated over $k$ to obtain the LDOS from an isolated (non-periodic) point source in the periodic structure~\cite{capolino2007comparison}. However, the effect of the EP is much clearer in the integrand (LDOS$_k$) than in the integral (LDOS), and so we focus here on the former for illustration purposes, exploiting the fact that $k$ allows us to control how close we are to an EP without altering losses.

Our example system consists of a waveguide  with periodic index modulation along its central axis, $\hat{x}$ [\figref{slab-passive}(a)].   PML  are used to truncate the transverse ($\hat{y}$) direction.
The design parameters are: $\varepsilon_1=12, \varepsilon_{2}=13.137$, $d=0.51a$, and $t=0.25a$. These parameters were chosen so that the corresponding one-dimensional system (with the same $\varepsilon_{1}, \varepsilon_2, d$ and infinite thickness,   $t\rightarrow\infty$) has nearly degenerate second and third frequency bands near $k_x=0$ [guaranteed by choosing parameters close to  the quarter-wave plate condition: $\sqrt{\varepsilon}_1d = \sqrt{\varepsilon}_2(a-d)$].
We force an EP for TM-polarized modes ($E_z, H_x, H_y$) by  fine-tuning the wavevector $k_x$ and the permittivity contrast $\delta\varepsilon\equiv\varepsilon_2-\varepsilon_1$ (see \appendixref{Appendix3}, \figref{threeD}).
\figrefbegin{slab-passive}(a) also depicts the real and imaginary parts of the coalescing eigenvalues (blue and red curves respectively). 
At $\vec{k}=0$ (also called the $\Gamma$ point), Maxwell's eigenvalue problem   is  complex-symmetric and, consequently, the   eigenvectors $\vec{E}_1^R$ and $\vec{E}_2^R$ are orthogonal (see mode profiles in the lower-left corner, having even/odd symmetry when $x$ is flipped to $-x$).
At the EP, the   eigenmodes $\vec{E}_{1,2}^R$ merge into  a single degenerate  mode $\vec{E}_0^R$ (upper-right panel). We also  compute the Jordan vector  ($\vec{J}^R_0$) with   our  novel iterative method~\cite{Hernandez2016}, and use it  to verify  our  formula for the  LDOS at the EP [\eqref{DoubleLDOS}].

 \subsection{Passive structures with EPs}
  \label{subsec:PassiveEP2}
 \figrefbegin{slab-passive}(b) depicts the LDOS$_k$ near the periodic  waveguide  at  several  wavevectors ($k_\mathrm{EP}, k_1, k_2, k_\infty$ marked on the lower plot).
 Far  from the EP (at  $k_\infty$), the LDOS$_k$ is a sum of two non-overlapping Lorentzian curves. As $k$ approaches the EP (e.g., at $k_1$ and  $k_2$), the resonance peaks  begin to overlap and the  LDOS$_k$ peaks  increase  due to the growing  Petermann factor. Most importantly, for $k$ values near but not equal to $k_\mathrm{EP}$, the traditional modal expansion formula [\eqref{GreenExpand}]  approaches the limiting Jordan-form-based formula [\eqref{DoubleGreen}].
Physically, this means that structures with nearby EPs can be approximated by truly defective structures, making this analysis useful for experimental systems, which  can only be close to but not exactly at an EP due to fabrication and calibration imperfections. Computationally, this implies that as long as Maxwell's operator $A(k)$ is not exactly defective, one can use \eqref{GreenExpand} to evaluate  the LDOS$_k$, when properly canceling the divergent terms  in   the sum.

The lower plot  in \figref{slab-passive}(b) compares the enhancement, $M_k/M_\infty$, for structures with varying quality factors, where we introduce the definitions  $M_k\equiv\gamma\cdot\max_\omega[\mathrm{LDOS}_k(\omega)]$ and $M_\infty\equiv\gamma\cdot\max_\omega[\mathrm{LDOS}_\infty(\omega)]$. We change $Q$ by modifying the permittivity contrast  $\delta\varepsilon$: Radiation losses decrease with decreasing index contrast of the periodic  modulation~\cite{benisty2000radiation}, with  the limit of zero  contrast corresponding to infinite $Q$. 
By plotting the normalized  LDOS$_k$ peak ($M_k/M_\infty$) vs.  deviation from the EP ($\Delta k\equiv k-k_\mathrm{EP}$), we find that the enhancement at the EP ($M_\mathrm{EP}/M_\infty$) is always   four-fold, regardless of $Q$. 
This follows from  a sum rule  which states that the spectrally integrated LDOS (and, therefore, also the LDOS$_k$) is a constant~\cite{barnett1996sum}. 
It implies that the maximum LDOS at an EP [i.e., the peak of a square Lorentzian $\frac{M_\mathrm{EP}\gamma^3}{[(\omega-\Omega_\mathrm{EP})^2+\gamma^2]^2}$] is four times larger than the maximum LDOS at a non-degenerate resonance [i.e., the peak of a simple Lorentzian $\frac{M_\infty\gamma}{(\omega-\Omega_\mathrm{EP})^2+\gamma^2}$].

\subsection{Passive structures with higher order EPs}
  \label{subsec:PassiveEP3}

Motivated by recent interest in higher-order EPs~\cite{graefe2008non,ryu2012analysis,heiss2015resonance,ding2016emergence}, 
we   generalize our results from the previous section to  $n$th order degeneracies  (i.e., EPs that form  when $n$ degenerate eigenvectors   merge). 
In this case, we define   the enhancement factor as the ratio of the LDOS peak at the EP and  at a reference point with $n$ non-degenerate resonances (generalizing our earlier definition for $M_\mathrm{EP}/M_\infty$).
Following  the arguments from \subsecref{simplifiedLDOS}, we expect to find a   squared Lorentzian emission curve at  second-order EPs,  a cubic Lorentzian  curve at  third-order EPs, and a Lorentzian to the $n$th power, $L_n(\omega) = \frac{M_n\gamma^{2n-1}}{[(\omega-\Omega_\mathrm{EP})^2+\gamma^2]^n}$, at $n$th order  EPs.
(This is  essentially equivalent to a known result on the rate of divergence of the norm of the resolvent matrix as an $n$th order EP is approached~\cite{trefethen2005spectra}.)
From the sum rule mentioned above~\cite{barnett1996sum},  the spectrally integrated LDOS
at an $n$th order EP, $S_n = \int d\omega L_n(\omega) = \frac{M_n\sqrt{\pi}\Gamma[n-\frac{1}{2}]}{\Gamma[n]}$,
is equal to the integrated LDOS before the $n$ resonances merge (here,  $\Gamma[n]$ is the gamma function). 
Realizing that the enhancement at the EP is maximal when merging $n$ identical resonances, a bound can be computed by solving  $S_n  = nS_1$. 
 We find that    the enhancement at  the EP is at most  $M_n/M_1 = \frac{\sqrt{\pi}\Gamma[n+1]}{\Gamma[n-\frac{1}{2}]}$. For example, at third order passive EPs ($n=3$), the enhancement  is at most 8-fold (as some of us recently confirmed  in~\cite{Lin2016}).

These results imply that higher order EPs could potentially provide a new route for achieving order-of-magnitude enhancement  of the  LDOS and order-of-magnitude narrowing of  the emission linewidth (in contrast to  traditional methods, which typically aim to 
 maximize the traditional Purcell factor by increasing the quality factor and reducing the mode volume ~\cite{Purcell1946}). 
However, this result does not necessarily mean that higher-order EPs will yield a larger   LDOS than the \emph{best} lower-order EP or single resonance. The reason is that the degrees of freedom that one would use to bring $n$ resonances together might otherwise be employed to enhance the $Q$ of an individual resonance. In practice, there will be a tradeoff between the quality of individual resonances and the order of the EP.

 \subsection{Active structures}
   \label{subsec:ActiveEP2}

\begin{figure}[t]
                 \centering\includegraphics[width=0.8\textwidth]{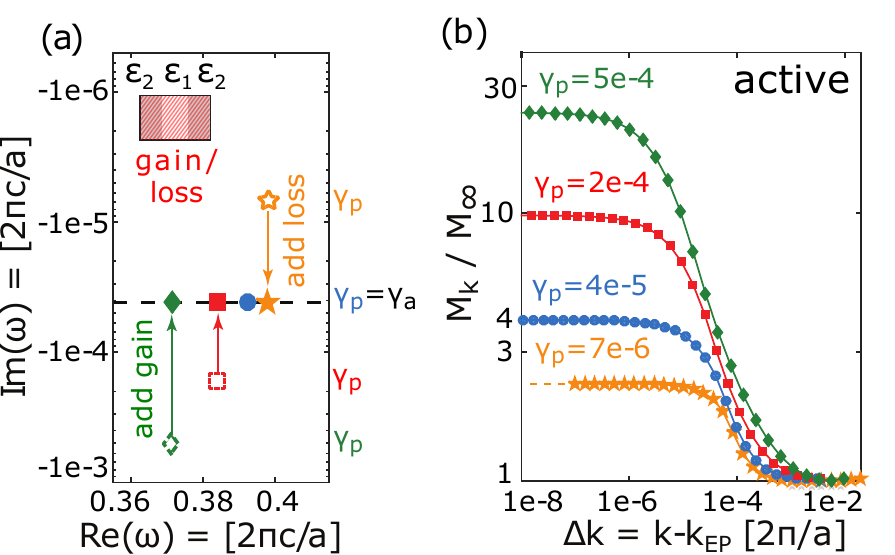}
                                   \caption{LDOS$_k$ enhancement in active structures.
 (a) Resonances with different passive loss rates $\gamma_{\text{p}}$ (hollow symbols) and equal active loss rates $\gamma_a$  (full symbols).  Inset:  Structure from \figref{slab-passive}(a) with   gain/loss added to  the waveguide.  (b) Normalized LDOS$_{k}$ peak  ($M_k/M_\infty$) vs. $\Delta k$, evaluated at $\vec{r}_0$ [marked in \figref{slab-passive}(a)], after adding gain/loss. The enhancement at the EP increases with the overall gain.
}
\label{fig:active-slab}
  \end{figure}

We show in this section that much greater     enhancements can be achieved in active systems, i.e., by  introducing  gain.  \figrefbegin{active-slab}  compares four periodic waveguides with different index contrasts ($\delta\varepsilon$), corresponding to different passive quality factors, $Q_p$. Gain and loss are added to each of the waveguides in order to force EPs, all of which share the same active quality factor $Q_a$.    
The empty (filled) markers  in \figref{active-slab}(a)  are  the EP  resonances ($\omega_\mathrm{EP}$) in the complex plane before (after) adding gain/loss. As shown in \figref{active-slab}(b), the structure with the smallest passive quality factor $Q_p$ 
has the largest LDOS$_k$  enhancement  at the EP since it requires more  gain in order to attain the same $Q_a$. 
Structures with  smaller initial $Q_p$ values can  lead to even greater enhancements, and we present such a case in \appendixref{Appendix4}. 
In principle, the relative enhancement at the EP, $M_\mathrm{EP}/M_\infty$, is not bounded in this computational model. However, in practice, giant \emph{relative} enhancements do not necessarily imply giant \emph{absolute} LDOS values (since  the value of the LDOS at the reference point may be very small).
Moreover, when adding enough gain to bring the system to the lasing threshold, stimulated emission eventually limits line narrowing and LDOS enhancement at the EP.

Similar to the analysis  of the plasmonic example in \secref{Plasmonics}, we can introduce a simplified  $2\times2$ model to interpret the results [analogous to \eqref{TCMTrestriction}]. We project the full Maxwell's operator $\hat{\mathcal{A}}$ onto the  subspace spanned by the two modes at $k_x = 0$ [shown in the left-lower corner of \figref{slab-passive}(a)] and we obtain
\begin{align}
\hat{A} = 
\left( \begin{array}{cc}
[\Omega_\mathrm{EP}\pm i\eta]^2 &  2\Omega_\mathrm{EP}v_gk\\
 2\Omega_\mathrm{EP}v_gk & [\Omega_\mathrm{EP}-2i\gamma_p\pm i\eta]^2
 \end{array} \right),
\end{align}
where  $v_g\equiv\left.\tfrac{\partial\omega}{\partial k}\right|_{k=0}$ is the the group velocity~(similar to the model used in \cite{zhen2015}). Prior to adding gain/loss, an EP occurs when $v_gk = \gamma_p$, and  the degenerate frequency is 
$\omega_\mathrm{EP}=\Omega_\mathrm{EP}-i\gamma_p$. After adding  gain/loss, $\pm i\eta$, the EP frequency  moves vertically in the complex plane, becoming $\Omega_\mathrm{EP} - i\gamma_a$, while the EP condition remains the same: $v_gk =\gamma_p$. In analogy to    \eqref{Qsqr_eq},   this model implies that the maximal enhancement at the EP scales as $M_\mathrm{EP} \propto \tfrac{\gamma_p}{\gamma_a^2}+\tfrac{1}{\gamma_a}$.
The quadratic scaling of $M_\mathrm{EP}$ with $Q_a$ in the limit of $\gamma_a\ll\gamma_p$ is demonstrated in \appendixref{Appendix4} [but this result is essentially another demonstration of the  spectral property shown in \figref{Qsqr}(b)].



 \section{Discussion}

The theory presented  in this work  provides a quantitative prescription for achieving large spontaneous emission rates  using  EPs, potentially exceeding by orders of magnitude those attained with standard non-degenerate resonances.   
Such enhancements  could  be  useful  for various applications, including fluorescent and Raman sensing~\cite{colthup2012introduction}, high-power low-coherence light sources~\cite{fujimoto2000optical}, and sources with tunable coherence~\cite{mckinney2000characterization}.
Moreover, by extending the current \emph{linear} theory  to account for EPs in nonlinear systems, our approach could  be applied to  study the properties of lasers at EPs---a  topic that has recently drawn great attention in the optics community ~\cite{brandstetter2014reversing,hodaei2014parity,feng2014single,miao2016orbital}. 

Our formulation for the LDOS  at the EP  [\eqsref{DoubleGreen}{DoubleLDOS}]  establishes that the enhancement  generally consists of two terms, one that scales linearly and one that scales quadratically with the quality factor. In \secsref{Plasmonics}{EnhancementLimits}, we  verified this scaling argument via two numerical examples,  and we employed simplified $2\times2$ models to estimate the coefficients of the quadratic terms in the LDOS formula. (In the plasmonic  system, we found that the coefficient was the coupling constant~$\kappa$, while in the periodic example,  the coefficient was the passive decay rate~$\gamma_p$.)  
More generally, we show in \appendixref{Appendix5} that   for arbitrary low-loss systems, the quadratic coefficient is bounded by
\begin{align}
\kappa \leq 
|\omega_\mathrm{EP}|\,\sqrt{\mathrm{max}\left|\mathrm{Im}\,\varepsilon\right|}.
\label{eq:perturbation}
\end{align}
This result provides an easy-to-evaluate  upper  bound on   the maximal   enhancement in  complicated geometries, depending only on the maximal gain/loss of the constituent materials.  
More explicitly, in active systems, we find that the maximal enhancement at the EP is bounded by 
$\tfrac{M_{\mathrm{EP}}}{M_\infty} = 
2(1 + \tfrac{\kappa}{\mathrm{Im}[\omega_\mathrm{EP}]}) 
\leq 2 
\left(1 + 2Q|\sqrt{\mathrm{max}\left|\mathrm{Im}\,\varepsilon\right|}\right)$. 
Note that in our plasmonic example, the quadratic coefficient is within $10\%$ of the bound.

Finally, our theory extends beyond spontaneous emission, and can be applied to a broader class of phenomena described by the LDOS. For example, we anticipate similar enhancements  in  higher-harmonic generation rates in nonlinear  media (e.g., Kerr media). In that case,  the input lower-harmonic field (multiplied by the nonlinear susceptibility) acts as a source to the higher-harmonic field.  To lowest order in the nonlinearity,  the converted power is found by  convolving the Green's function with the input signal. This is analogous to our formulation of spontaneous emission (which involved  convolving a dipole source with the Green's function), and would therefore result in similar enhancements~\cite{Pick2016}. More generally, a similar treatment could produce enhancements in related quantities at EPs in other areas of physics (e.g., exciton-polariton~\cite{gao2015observation}  mechanical systems~\cite{xu2016topological}, and also leak-wave antennas both at radio-frequency and visible frequencies~\cite{jackson2008leaky,monticone2015leaky}). Finally,  our theory can be generalized to study scattering and extinction problems. We find, however, that even though EPs can produce special spectral features in scattering cross-sections, they do not give rise to giant scattering enhancements,  since the scattered intensity is bounded by the incoming intensity (and the bound of perfect scattering can be easily achieved with a single non-degenerate resonance~\cite{miller2016fundamental}).


\appendix

\section*{Appendix}

\appendixsec{Non-degenerate Green's function modal expansion formula}{Appendix1}

In this section we review the derivation of the \eqref{GreenExpand} in the main text. 
Our derivation  is similar to standard methods for \emph{Hermitian} eigenvalue problems~\cite{Arfken2006}, with the necessary modifications for treating general \emph{non-Hermitian} systems (most importantly, using the unconjugated ``inner product''~\cite{Moiseyev2011} between left and right eigenmodes).

Given a non-magnetic medium with dielectric permittivity $\varepsilon$, the   
fields $\vec{E}$ and currents  $\vec{J}$ in the medium are related via Maxwell's frequency-domain partial differential equation:
\begin{equation}
(\hat{\mathcal{A}}_\vec{r}-\omega^2\unit) \vec{E}(\vec{r})=i\omega\vec{J}(\vec{r}).
\end{equation}
where  $\hat{\mathcal{A}}_\vec{r}\equiv\frac{1}{\varepsilon}\nabla_\vec{r}\times\nabla_\vec{r}\times$. 
The response to arbitrary currents can be found by  
convolving the dyadic Green's function with the current sources: $\vec{E}(\vec{r}) = \int d\vec{r}'\mat{G}(\vec{r},\vec{r}')\vec{J}(\vec{r}')$, where $\mat{G}$ is defined via:
\begin{equation}
(\hat{\mathcal{A}}_{\vec{r}}-\omega^2\unit) \mat{G}(\vec{r},\vec{r}') = -\delta(\vec{r}-\vec{r}')\unit.
\label{eq:GreenPDE}
\end{equation}

In this section, we expand $G$ in terms of   right and left  resonant modes ($\vec{E}^R_n$  and $\vec{E}^L_n$), which  are outgoing solutions of the partial differential equations
\begin{align}
&(\hat{\mathcal{A}}_\vec{r}-\omega_n^2\unit) \vec{E}^R_n(\vec{r})= 0\nonumber\\
&(\hat{\mathcal{A}}^T_\vec{r}-\omega_n^2\unit) \vec{E}^L_n(\vec{r})= 0.
\label{eq:LeftRieghtPDEs}
\end{align}
When the set of eigenvectors of  \eqref{LeftRieghtPDEs} forms a complete basis of the Hilbert space, 
one can  introduce the completeness relation, which consists of expanding the Dirac delta function as
\begin{equation}
\delta(\vec{r}-\vec{r}')\unit=
\sum_n
\vec{E}_n^R(\vec{r})\otimes \vec{E}^L_n(\vec{r}'),
\label{eq:deltaExpand}
\end{equation}
where $\otimes$ is the outer/tensor product $\vec{u} \otimes \vec{v} = \vec{u} \vec{v}^T$.
The  question of  completeness of eigenmodes in non-Hermitian open systems has not been proven in general, for arbitrary three-dimensional systems. However, since in this work we always evaluate the Green's function in close proximity to the resonators and at frequencies close to the resonance frequencies, the  eigenmodes which overlap spectrally and spatially with the emitter give a good approximation for the LDOS, justifying the use of \eqref{deltaExpand}.

Similarly, we wish to find an  expansion formula for  $\mat{G}$ or, more explicitly, find the coefficients $\vec{a}_n(\vec{r}')$ in the series
\begin{equation}
\mat{G}(\vec{r},\vec{r}')=
\sum_n\vec{E}^R_n(\vec{r})\otimes \vec{a}_n(\vec{r}').
\label{eq:expand}
\end{equation}
To this end, we substitute   \eqref{deltaExpand} and  \eqref{expand} into \eqref{GreenPDE} and obtain
\begin{equation}
\sum_n \left(\mathcal{A}_{\vec{r}}-\omega^2\right)\vec{E}^R_n(\vec{r})\otimes \vec{a}_n(\vec{r}')=
-\sum_n\vec{E}^R_n(\vec{r})\otimes \vec{E}^L_n(\vec{r}').
\end{equation}
Next, we multiply both sides of the equation by ${[\vec{E}^L_m(\vec{r})]}^T$. Using the relation: $[\vec{E}^L_m(\vec{r})]^T\mathcal{A}_{\vec{r}} = \omega_m^2[\vec{E}^L_m(\vec{r})]^T$,
integrating  over $\vec{r}$, and invoking the bi-orthogonality relation for non-Hermitian systems~\cite{Siegman2000Frontiers}:
\begin{equation}
\int\!dr\,\vec{E}^L_m(\vec{r})\cdot \vec{E}^R_n(\vec{r}) = \delta_{m,n},
\end{equation}
we obtain 
$\vec{a}_m(\vec{r}') = 
\frac{\vec{E}^L_m(\vec{r}')}{\omega^2-\omega_m^2}$. Finally,  substituting this result in \eqref{expand} we obtain an eigenmode expansion of the dyadic Green's function [which reduces to  \eqref{GreenExpand} in the text]:
\begin{equation}
\mat{G}(\vec{r},\vec{r}',\omega)=\sum_n 
\frac{1}{\omega^2-\omega_n^2}
\frac{\vec{E}^R_n(\vec{r})\otimes \vec{E}^L_n(\vec{r}')}
{\int\!dr\,\vec{E}^R_n(\vec{r})\cdot\vec{E}^L_n(\vec{r})}.
\label{eq:simple}
\end{equation}
The integral in the denominator of \eqref{simple} is one, but we keep it here for comparison with \eqref{GreenExpand}.

Last, we note that in reciprocal media $\varepsilon = \varepsilon^T$, and  there exists a simple relation between left and right eigenvectors: $\vec{E}_n^L = \varepsilon \vec{E}_n^R$. More generally, the left and right eigenvectors of a symmetric generalized eigenvalue problem (EVP): $\mat{A}\vec{E}_n^R = \lambda_n \mat{B} \,\vec{E}_n^R, 
\mat{A}^T\vec{E}_n^L = \lambda_n \mat{B}^T \vec{E}_n^L$, with $A = A^T$ and $B = B^T$ are related via $\vec{E}_n^L = \mat{B}\vec{E}_n^R$. To see this, rewrite the EVP for the left vectors as 
$(\mat{B}^{-1}\mat{A})^T \vec{E}_n^L = \mat{A} (\mat{B}^{-1} \vec{E}_n^L) = \lambda_n \mat{B}(\mat{B}^{-1}\vec{E}_n^L)$ which shows that $\mat{B}^{-1}\vec{E}_n^L$ is a right eigenvector. Note that this relation holds also with  Bloch-periodic boundary conditions (and the surface-term correction found in~\cite{vial2014quasimodal} does not appear in our formulation). Although the matrix $\mat{A}$ is no longer symmetric in the $k$-periodic problem, it satisfies the relation $\mat{A}(k)^T = \mat{A}(-k)$, and since we are essentially relating $k$-right eigenvectors to $(-k)$-left eigenvectors, the relation above remains unchanged.

\appendixsec{The effect of dispersion on the LDOS formula}{AppendixNew}

In this appendix, we consider the effects of dispersion on the Green's function near and at the EP. In accordance  with previous work on quasi-normal modes in dispersive media~\cite{perrin2016eigen,sauvan2013theory}, we find that  the Green's function has non-diagonal contributions $\propto \vec{E}_\pm^L\otimes\vec{E}_\mp^R$ near the EP. However, exactly at the EP,   the Green's function has exactly the same form as the non-dispersive degenerate formula [\eqref{DoubleGreen}], with dispersion affecting only the normalization of the degenerate mode ($\vec{E}_0$) and Jordan vector ($\vec{J}_0$). 

In the same spirit as our derivation  of the non-dispersive formula, we expand the Green's function $\mat{G}$ in eigenmodes:
\begin{align}
\mat{G}(\vec{r},\vec{r}_0,\omega)  = \sum_m \vec{E}^R_m(\vec{r})\otimes\vecg{\alpha}_m(\omega,\vec{r}_0),
\label{eq:GreenDispersionsExpand}
\end{align}
and   use  simple algebraic manipulations to express the coefficients $\vecg{\alpha}_m$ in terms of the modes. 
Recall that $\mat{G}$ is defined as the solution to the partial differential equation:
\begin{equation}
\left[\nabla\times\nabla \times- \omega^2\varepsilon(\vec{r},\omega)\right]\mat{G}(\vec{r},\vec{r}_0,\omega) = -\delta(\vec{r}-\vec{r}_0)\unit
\label{eq:GreenDefinition}
\end{equation}
By  multiplying both sides of \eqref{GreenDefinition}  from the left
 by $\vec{E}_n^L(\vec{r})$ and integrating over $dr$, we obtain
\begin{equation}
\vec{E}^L_n(\vec{r}_0) = 
\int\!dr\, \vec{E}^L_n(\vec{r})\left[\omega^2\varepsilon(\vec{r},\omega) - \omega_n^2\varepsilon(\vec{r},\omega_n) \right]
\mat{G}(\vec{r},\vec{r}_0,\omega) 
\label{eq:GeneralizedIP}
\end{equation}
Then, by substituting \eqref{GreenDispersionsExpand} into \eqref{GeneralizedIP}, we find
\begin{align}
& \vec{E}_n^L(\vec{r}_0) = \nonumber\\
&\sum_m 
\int\! dr\,
\vec{E}_n^L(\vec{r})
[\omega^2\varepsilon(\omega)-\omega_n^2\varepsilon(\omega_n)]\vec{E}_m^R(\vec{r}) \otimes \vecg{\alpha}_m(\omega,\vec{r}_0)
\equiv\sum_m  (\vec{E}_m,\vec{E}_n)\vecg{\alpha}_m(\omega,\vec{r}_0).
\label{eq:ModeAtx0}
\end{align}
(Note that this result was also  derived in~\cite{sauvan2013theory} by invoking Lorentz reciprocity.)
Since we are interested in  emission from emitters in close proximity to the resonators and in frequencies near the resonant frequencies, we may assume that  a finite set of $N$ eigenvectors adequately describes the system's response [similar to our assumption in the non-dispersive derivation in  \eqref{deltaExpand}]. We introduce the vector $\vec{s}(\vec{r}) = \{\vec{E}_1(\vec{r}),\hdots,\vec{E}_N(\vec{r})\}$ and the matrix $\mat{O}_{mn}\equiv (\vec{E}_m,\vec{E}_n) $, and rewrite \eqref{ModeAtx0}  as: $\vec{s}(\vec{r}_0) = \mat{O}(\omega)\vecg{\alpha}(\omega,\vec{r}_0)$ or equivalently 
$\vecg{\alpha}(\omega,\vec{r}_0) = \mat{O}^{-1}(\omega)\vec{s}(\vec{r}_0)$.
With this notation, \eqref{GreenDispersionsExpand} can be rewritten as: $\mat{G} = \vecg{\alpha}(\omega,\vec{r}_0)^T\vec{s}(\vec{r}) = \vec{s}(\vec{r}_0)^T {(\mat{O}^{-1})}^T\vec{s}(\vec{r}_0)$ or 
\begin{align}
\mat{G}(\vec{r},\vec{r}_0,\omega) =  
 \sum_{nm}
 ({\mat{O}^{-1}})_{nm}
 \vec{E}_n^R(\vec{r})\otimes  
  \vec{E}_m^L(\vec{r}_0)
\label{eq:NewGreen}
\end{align}
where
\begin{align}
\mat{O}_{ij} = \int \!dr\, \vec{E}_i^L [\omega^2\varepsilon(\vec{r},\omega)-\omega_i^2\varepsilon(\vec{r},\omega_i)]\vec{E}^R_j.
\label{eq:Omat}
\end{align}
\eqrefbegin{NewGreen} and \eqref{Omat} are the main result of this appendix---an eigenmode expansion of the Green's function in the presence of dispersion, generalizing \eqref{GreenExpand} from the main text. 
In the non-dispersive limit, this expression reproduces our previous result since $\mat{O}_{mn}  = (\vec{E}_m,\vec{E} _n) = (\omega^2 - \omega_n^2)\int dr\,  \varepsilon\,\vec{E}^L_m \cdot \vec{E}^R_n = (\omega^2 - \omega_n^2)\delta_{mn}$ and 
$\mat{G}= \sum_n(\omega^2 - \omega_n^2)^{-1}\vec{E}^R_n(\vec{r})\otimes \vec{E}^L_n(\vec{r}_0)$.

Next, let us calculate the limit of \eqref{NewGreen} when two modes $\vec{E}_\pm$ coalesce at an EP. 
Keeping  only the terms corresponding to $\vec{E}_\pm$ in \eqref{NewGreen}, 
expanding  both $\omega$ and $\omega_\pm$ in Taylor series around $\omega_0$ and introducing the notation:
\begin{align}
\omega^2\varepsilon(\omega) - \omega_\pm^2\varepsilon(\omega_\pm) = 
(\omega-\omega_\pm)
\left.
\tfrac{d(\omega^2\varepsilon)}{d\omega}
\right|_{\omega_0}\equiv
(\omega-\omega_\pm)(\omega^2\varepsilon)'_0\,,
\end{align}
we find  that $\mat{O}_{ij} = (\omega - \omega_j)\int dr (\omega^2\varepsilon)'_0\vec{E}_i^L\cdot\vec{E}_j^R $, and the matrix inverse is
\begin{align}
\mat{O}^{-1} = 
\left( \begin{array}{cc}
\tfrac{1}{\omega-\omega_+}
\tfrac{\int  (\omega^2\varepsilon)'_0\vec{E}_-^L\cdot\vec{E}_-^R}{N}
 & 
 -\tfrac{1}{\omega-\omega_+}
\tfrac{\int  (\omega^2\varepsilon)'_0\vec{E}_+^L\cdot\vec{E}_-^R}{N}
  \\
 -\tfrac{1}{\omega-\omega_-}
\tfrac{\int  (\omega^2\varepsilon)'_0\vec{E}_-^L\cdot\vec{E}_+^R}{N}
 & \tfrac{1}{\omega-\omega_-}
\tfrac{\int  (\omega^2\varepsilon)'_0\vec{E}_+^L\cdot\vec{E}_+^R}{N}
 \end{array} \right),
 \label{eq:stepGreen}
\end{align}
where  $N\!\equiv \!
\left[\int  (\omega^2\varepsilon)'_0\vec{E}_+^L\cdot\vec{E}_+^R\right]
\left[\int  (\omega^2\varepsilon)'_0\vec{E}_-^L\cdot\vec{E}_-^R\right]-
\left[\int  (\omega^2\varepsilon)'_0\vec{E}_+^L\cdot\vec{E}_-^R\right]
\left[\int  (\omega^2\varepsilon)'_0\vec{E}_-^L\cdot\vec{E}_+^R\right]$. Substituting \eqref{stepGreen} into \eqref{NewGreen}, we find that   the Green's function near the EP is
\begin{align}
\mat{G}(\vec{r},\vec{r}_0,\omega) &= 
\tfrac{\vec{E}_+^R\otimes \vec{E}_+^L}{\omega-\omega_+}
\tfrac{\int  (\omega^2\varepsilon)'_0\vec{E}_-^L\cdot\vec{E}_-^R}{N}+
\tfrac{\vec{E}_-^R\otimes \vec{E}_-^L}{\omega-\omega_-}
\tfrac{\int  (\omega^2\varepsilon)'_0\vec{E}_+^L\cdot\vec{E}_+^R}{N}-\nonumber\\
&\tfrac{\vec{E}_-^R\otimes \vec{E}_+^L}{\omega-\omega_+}
\tfrac{\int  (\omega^2\varepsilon)'_0\vec{E}_+^L\cdot\vec{E}_-^R}{N}
-\tfrac{\vec{E}_+^R\otimes \vec{E}_-^L}{\omega-\omega_-}
\tfrac{\int  (\omega^2\varepsilon)'_0\vec{E}_-^L\cdot\vec{E}_+^R}{N} .
\label{eq:GreenStepStep}
\end{align}
Last, we want to take the limit of \eqref{GreenStepStep} as the two modes $\vec{E}_\pm$ approach the EP. We expand $\omega_\pm$ and $\vec{E}_\pm$ around the EP using degenerate perturbation theory (using the notation  of \subsecref{DegenPert}):
\begin{align}
\omega_\pm \approx \omega_0 \pm \sqrt{p}\Delta ,\nonumber\\
\vec{E}_\pm \approx \vec{E} _0 \pm \sqrt{p}\Delta\,\vec{J}_0.
\end{align}
Generalizing  our derivation of the non-dispersive formula, we choose the normalization conditions
$\int (\omega^2\varepsilon)'_0\vec{E}_0^L\vec{J}_0^R  = \int (\omega^2\varepsilon)'_0\vec{J}_0^L\vec{E}_0^R = 1$ and  $\int (\omega^2\varepsilon)'_0\vec{J}_0^L\vec{J}_0^R = 0$, while the condition $\int (\omega^2\varepsilon)'_0\vec{E}_0^L\vec{E}_0^R = 0$ is automatically satisfied   due to self-orhotognality. With this normalization, we can calculate  
 integrals in \eqref{GreenStepStep}: 
\begin{align*}
\int (\omega^2\varepsilon)'_0\vec{E}_\pm^L\vec{E}_\pm^R&= 
\cancel{\int (\omega^2\varepsilon)'_0\vec{E}_0^L\vec{E}_0^R} \pm
2\sqrt{p}\Delta\int (\omega^2\varepsilon)'_0\vec{E}_0^L\vec{J}_0^R  + 
p\Delta^2\cancel{\int (\omega^2\varepsilon)'_0\vec{J}_0^L\vec{J}_0^R}+\mathcal{O}(p^{3/2})
 \nonumber\\
\int (\omega^2\varepsilon)'_0\vec{E}_\pm^L\vec{E}_\mp^R&= 
\cancel{\int (\omega^2\varepsilon)'_0\vec{E}_0^L\vec{E}_0^R} - p\Delta^2
\cancel{\int (\omega^2\varepsilon)'_0\vec{J}_0^L\vec{J}_0^R}+\mathcal{O}(p^{3/2})\nonumber \\
&N = -4p\Delta^2 \int (\omega^2\varepsilon)'_0\vec{E}_0^L\vec{J}_0^R+\mathcal{O}(p^{3/2})
\end{align*}
The first two terms in \eqref{GreenStepStep} have the same form as the non-dispersive expansion [\eqref{Limit}]: 
$\frac{1}{2\Delta\sqrt{p}}
\frac{\vec{E}_+^R\otimes \vec{E}_+^L}{\omega-\omega_+}-\frac{1}{2\Delta\sqrt{p}}
\frac{\vec{E}_-^R\otimes \vec{E}_-^L}{\omega-\omega_-} + \mathcal{O}(p^{3/2})$. Upon approaching the EP, each contributions diverges with an opposite sign, and a finite contribution remains. The last two terms scale as $\sqrt{p}$ near the EP, and vanish at the EP. Finally, we obtain 
\begin{align}
\mat{G}_\mathrm{EP}(\vec{r},\vec{r}_0,\omega) = 
\frac{\vec{E}_0^R(\vec{r})\otimes \vec{E}_0^L(\vec{r}_0)}{(\omega-\omega_0)^2}
+
\frac{\vec{J}_0^R(\vec{r})\otimes \vec{E}_0^L(\vec{r}_0)+
\vec{E}_0^R(\vec{r})\otimes \vec{J}_0^L(\vec{r}_0)}{\omega-\omega_0}.
\label{eq:GreenDispersionResult}
\end{align}
\eqrefbegin{GreenDispersionResult}  implies that the green's function at the EP in dispersive media has the same form as the non-dispersive formula [\eqref{DoubleGreen} in the text], but the normalization of the modes changes.

\appendixsec{Convergence of the ``unconjugated norm''}{Appendix6}

In this appendix, we show that when perfectly matched layers (PML) are used to implement outgoing boundary conditions, the ``unconjugated norm'' of a scattering eigenmode, $\int\!dx\,\varepsilon(\vec{x})\vec{E}_n(\vec{x})^2$,  converges to a finite result  as the PML thickness tends to infinity. 
Scattering eigenmodes are solutions to Maxwell's eigenvalue problem, $\varepsilon^{-1}\nabla\times\nabla\times\vec{E}_{n}=\omega_{n}^2 \vec{E}_{n}$~\cite{joannopoulos2011photonic}, with  outgoing radiation conditions.  These solutions (also called ``leaky modes''~\cite{hu2009understanding}) have complex frequencies $\omega_n$ which lie in the lower-half of the complex plane ($\mbox{Im}[\omega_n]<0$)~\cite{collin1960field} and, consequently, the  modal amplitudes ($|\vec{E}_n|$) grow unboundedly  at large distances from the structure.
Even though the modal amplitude diverges, we show that the unconjugated inner product is finite and independent of the PML parameters, as long as the PML works effectively (i.e., designed so that outgoing waves are attenuated exponentially inside the PML and are not reflected at the air-PML interface.) We provide a simple proof for  a one dimensional geometry.  An  abstract proof for the independence of the unconjugated norm on the  PML parameters appeared in  \cite{sauvan2013theory} (using a generalized definition of the norm, which includes dispersion).
An alternative one-dimensional proof was given in~\cite{leung1994completeness}, where the authors used analytic continuation of the coordinates, similar to the  coordinate stretching method that we use here.

\begin{figure}[b]
                 \centering\includegraphics[width=0.8\textwidth]{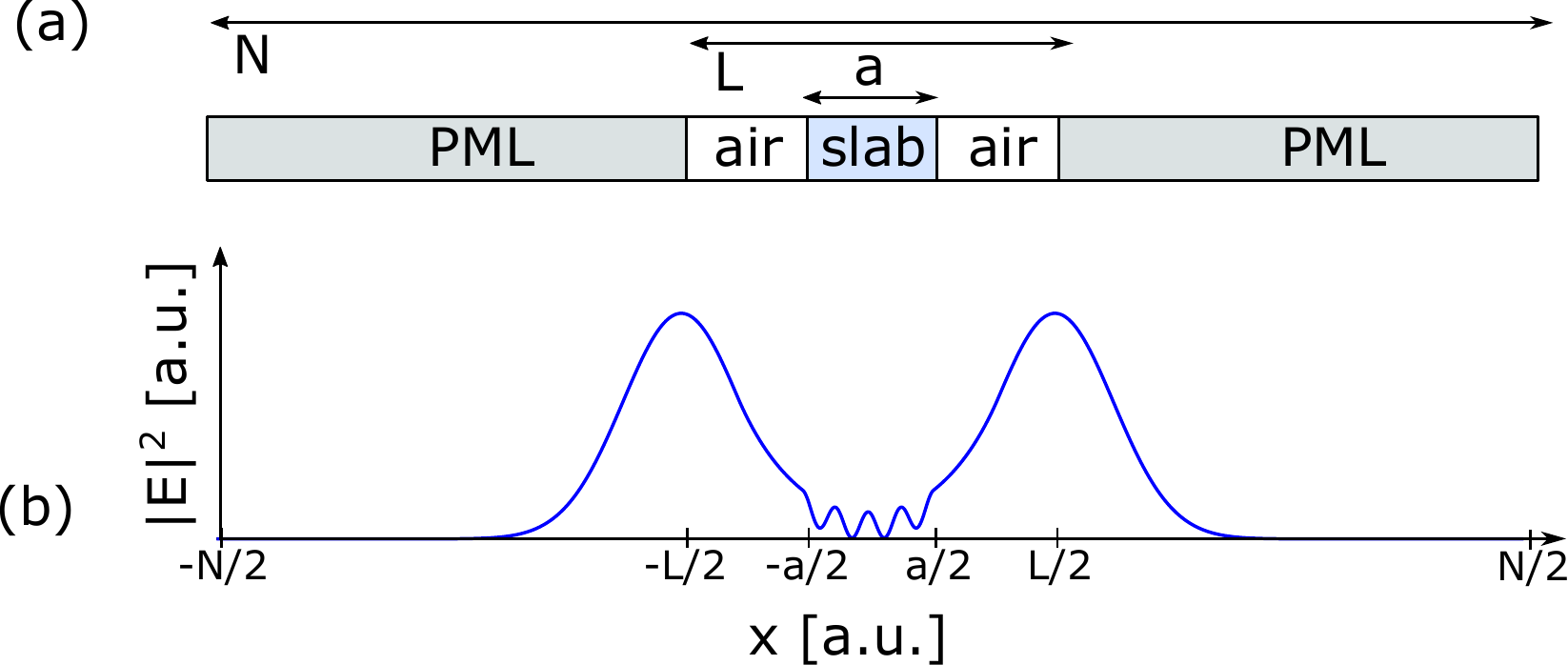}
                                   \caption{Leaky modes in one dimension.   
(a) A slab in air with PML at the cell boundary.  
(b) The  intensity of the leaky mode  increases exponentially in the air region  and is attenuated in  the PML.}
\label{fig:PMLconverge}
  \end{figure}

We consider the  one-dimensional geometry depicted in \figref{PMLconverge}(a), which consists of  a slab of  thickness $a$ and refractive index $n$, embedded in air. (Generalizations to more complex one-dimensional structures straightforwardly follow.)
We truncate the computational cell by placing PML  at a finite distance from the slab ($|x|= L/2$), and we impose metallic boundary conditions at the cell boundary ($x=\pm\tfrac{N}{2}$). 
The PML can be viewed as an analytic continuation of the spatial coordinate into the complex plane~\cite{chew19943d}, $\widetilde{x}=x+if(x)$, where the derivative of $f$ satisfies $\frac{df}{dx}=\frac{\sigma(x)}{\omega}$. (The latter condition guaranties that fields oscillating at different frequencies $\omega$ will be attenuated inside the PML at rate that is independent of $\omega$.)  For concreteness, we choose: $\sigma(x) = \frac{\sigma_0 x}{d}$, where $d=\tfrac{N-L}{2}$ is the PML thickness. 
Consequently,  the coordinate stretching transformation inside the PML is 
\begin{align}
\widetilde{x}=x+
i\frac{\sigma_0\left(x-\tfrac{L}{2}\right)^2}{2\omega d}.
\end{align}

Scattering solution  for this geometry can be written explicitly  as 
\begin{equation*}
E_n=
\left\{ \begin{array}{cc}
A_ne^{-i\omega_n x-({\sigma_0}/{2d})\left(x+L/2\right)^2}+
B_ne^{i\omega_n x+({\sigma_0}/{2d})\left(x+L/2\right)^2} & 
-N/2 < x < -L/2\\
A_ne^{-i\omega_n x}+B_ne^{i\omega_n x} & 
-L/2 < x < -a/2\\
e^{-i\omega_n nx}+e^{i\omega_n nx} & 
-a/2 < x < a/2\\
A_ne^{i\omega_n x}+B_ne^{-i\omega_n x} & 
a/2 < x < L/2\\
A_ne^{i\omega_n x-({\sigma_0}/{2d})\left(x-L/2\right)^2}+
B_ne^{-i\omega_n x+({\sigma_0}/{2d})\left(x-L/2\right)^2} & 
L/2 < x < N/2
 \end{array} \right.,
\label{eq:PTfield}
\end{equation*}
where the coefficients $A_n, B_n$ and resonant frequency $\omega_n$ are found by requiring continuity of the field ($E_n$) and its derivative 
($dE_n/dx$) at the interfaces, while imposing the boundary condition:  
$E_n=0$ at $\pm N/2$. We obtain 
\begin{align*}
A_n &= \frac{e^{inka/2}+e^{-inka/2}}
{e^{ika/2}-C_n e^{-ika/2}
},
\nonumber\\
B_n &=-A_nC_n\,,
\end{align*}
where we introduced $C_n\equiv e^{i\omega_n N-\sigma_0d}$.

Next, we compute the unconjugated  norm, $\int\varepsilon E_n^2$, and study its convergence.
Introducing the antiderivative function
\begin{align}
J(x)\equiv
\int^{x}\!dx'\,\varepsilon(A_ne^{i\omega_n x'}+B_ne^{-i\omega_n x'})^2,
\label{eq:A_nntiderivative}
\end{align}
the unconjugated norm can be written as
\begin{align}
&\mathcal{I}(d)\equiv
\int_\mathrm{all}\!dx\,\varepsilon E^2 =
\int_\mathrm{Cav}\!dx\,\varepsilon E^2+2\left[
J\left(\tfrac{L}{2}\right)-
J\left(\tfrac{a}{2}\right)
\right]+
2\left\{
J\left[\widetilde{x}(\tfrac{N}{2})\right]-
J\left[\widetilde{x}(\tfrac{L}{2})\right]
\right\}.
\label{eq:Unconjugated}
\end{align}
We now show that whenever the condition \eqref{condition} (below) holds,
$\mathcal{I}$ is finite and independent of the PML parameters in the limit of $d\rightarrow\infty$. 
Since the coordinate stretching factor is zero at the air-PML boundary [$\widetilde{x}(\tfrac{L}{2})=\tfrac{L}{2}$], the antiderivative terms at $L/2$ cancel ($J\left(\tfrac{L}{2}\right)=J\left[\widetilde{x}(\tfrac{L}{2})\right]$).   
Next,  consider   the boundary  term, $J\left[\widetilde{x}(\tfrac{N}{2})\right]$. Straightforward integration of \eqref{A_nntiderivative} yields three terms, all of which decay exponentially as $d\rightarrow\infty$, provided that $|C_n|$ decays exponentially. 
[The first is $\frac{A_n^2e^{2i\omega_n {\widetilde{x}(N/2)}}}{2i\omega_n }\!\!=\!\!\tfrac{A_n^2C_n}{2i\omega_n }$, the second term is $\frac{B_n^2e^{-2i\omega_n {\widetilde{x}(N/2)}}}{-2i\omega_n }\!\!=\!\!\tfrac{A_n^2C_n}{-2i\omega_n }$, and the last term is $2A_nB_n{\widetilde{x}(N/2)}=-2A_n^2C_n\cdot\mathcal{O}(d)$.] 
Introducing   $\omega_n =\omega_n '-i\omega_n ''$ and $\alpha = d/L$,
one finds from the definition of $C_n$ that it  decays exponentially whenever 
\begin{align}
[\omega_n ''(\alpha+2)+\sigma_0]d\gg1.
\label{eq:condition}
\end{align}
Evaluating the remaining terms in \eqref{Unconjugated}, $\int_\mathrm{Cav}\!dx\,\varepsilon E^2$ and $J\left(\tfrac{a}{2}\right)$,  we obtain:
\begin{align*}
\lim_{d\rightarrow\infty}
\mathcal{I}(d)=
\tfrac{2}{n\omega_n }\left[
\sin(n\omega_n a )+n\omega_n a 
\right]
+\frac{4i}{\omega_n }\cos\left(
\tfrac{n\omega_n a }{2}
\right).
\end{align*} 
The result is finite, and independent of the location of the air-PML interface or the cell boundary, thus completing the proof.

\begin{figure}[t]
                 \includegraphics[width=1\textwidth]{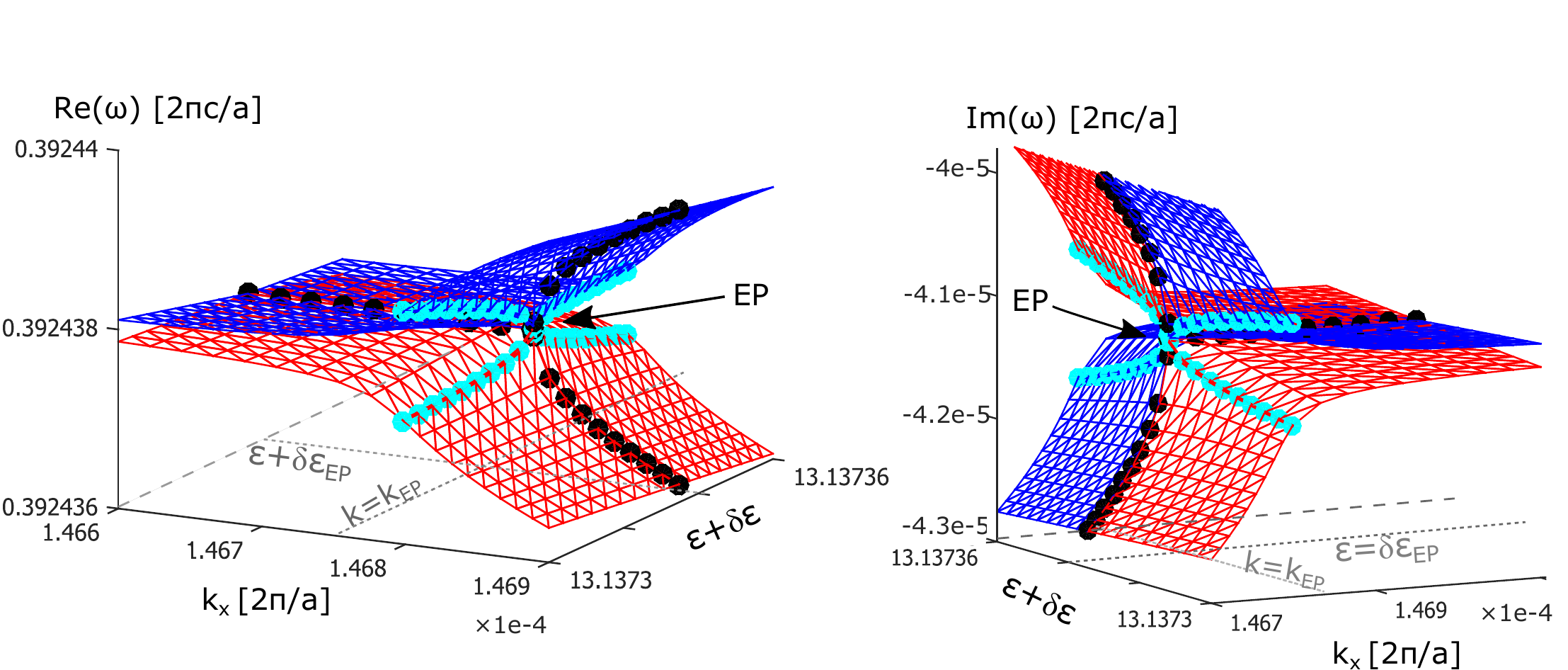}
                                   \caption{
                                   Forcing EPs in the periodic waveguides. (a) Real and (b) imaginary parts of the  eigenvalues of Maxwell's operator, as a function of the wavevector $k_x$ and the permittivity contrast  $\delta\varepsilon\equiv\varepsilon_2-\varepsilon_1$, for the geometry from  \figref{slab-passive}(a). The EP  is found by numerically minimizing the distance between the two eigenvalue sheets (red and blue surfaces). Cyan dots: Eigenvalues for fixed $k_x = k_\mathrm{EP}$ and  varying $\delta\varepsilon$. Black dots: Eigenvalues for fixed $\delta\varepsilon = \delta\varepsilon_\mathrm{EP}$ and   varying $k_x$.}
\label{fig:threeD}
  \end{figure}

\appendixsec{Computational details on Forcing  EPs in periodic waveguides}{Appendix3}

In order to force an EP (i.e.,  an accidental  degeneracy of two nearby resonances), we need to satisfy a single complex equation ($\omega_m = \omega_n$), which can be done by searching two  real parameters. 
In the plasmonic example in \secref{Plasmonics}, we search the 
two-dimensional  parameter space spanned by the gain/loss parameter ($|\mathrm{Im}\,\varepsilon|$) and the (real part) of the refraction index in the upper-half of the silica coating.  
In the periodic example from \secref{EnhancementLimits}, we search the two-dimensional  space spanned by the wavevector $k_x$ and the permittivity contrast $\delta\varepsilon$.  
 \figrefbegin{threeD} presents our numerical results for the periodic example. As shown,  we find a degenerate resonance at  $\omega_\mathrm{EP}=0.3924377 - 0.00004119303i$ when 
$k_\mathrm{EP}a/2\pi \approx 1.468\times10^{-4}$ and $\delta\varepsilon_\mathrm{EP} \approx 1.137$.

\appendixsec{Derivation of the simplified model for the LDOS}{Appendix2}

In this section, we derive \eqref{TCMTrestriction} from the main text, which provides a simplified formula for the LDOS in \secref{Plasmonics}.  
Our approach  is similar to  coupled-mode theory, originally developed  for  photonic  waveguides~\cite{huang1994coupled,okamoto2010fundamentals}. 
Our derivation consists of projecting   Maxwell's operator [describing  the  coupled-rod system from    \figref{Structure}(a)] onto the subspace spanned by the modes of the corresponding  uncoupled-system (i.e., a system  in which    the rod--rod separation  is infinite). 

The field of the (original)  coupled system  satisfies  Maxwell's  equation:
\begin{align}
\hat{\mathcal{A}}\,\vec{E} \equiv
\tfrac{1}{\varepsilon}\nabla\times\nabla\times \vec{E} = \omega^2 \vec{E}.
\label{eq:MaxwelEq}
\end{align}
Denoting by $\varepsilon_{\ell}$ the permittivity of a system that contains only a single  rod [where  $\ell=1$ ($2$)  indicates  the upper (lower) rod in \figref{Structure}(a)],
the right  eigenvectors  of the uncoupled system satisfy 
\begin{align}
\tfrac{1}{\varepsilon_{\ell}}\nabla\times\nabla\times \vec{E}_\ell^R = \omega_\ell^2\vec{E}_\ell^R,
\label{eq:uncoupled}
\end{align}
with a similar definition for  left eigenvectors. 

Next, let us compute  the projection of $\mathcal{A}$ onto the subspace spanned by $\vec{E}_1$ and $\vec{E}_2$, denoted by $\hat{A}$.
The diagonal terms of $\hat{A}$ are 
\begin{align}
\int\vec{E}_j^L\tfrac{1}{\varepsilon}\nabla\times\nabla\times\vec{E}_j^R\approx
\int\vec{E}_j^L\tfrac{1}{\varepsilon_j}\nabla\times\nabla\times\vec{E}_j^R = \omega_j^2.
\end{align}
The first approximation follows from the fact that  the mode profile $\vec{E}_j^R$ is significant only in close proximity to resonator $j$, and in that region  $\varepsilon = \varepsilon_j$. The second equality follows from \eqref{uncoupled}.
Now, let us define  $\Delta\varepsilon_j\equiv\varepsilon-\varepsilon_j$. The off-diagonal terms of $\hat{A}$ are 
\begin{align}
&\int\vec{E}^L_i\tfrac{1}{\varepsilon}\nabla\times\nabla\times\vec{E}^R_j=
\int\vec{E}^L_i\tfrac{\varepsilon_j}{\varepsilon}\tfrac{1}{\varepsilon_j}\nabla\times\nabla\times\vec{E}^R_j=\nonumber\\
&\omega_j^2\int\vec{E}^L_i\tfrac{\varepsilon-\Delta\varepsilon_j}{\varepsilon}\vec{E}^R_j
 = 2\omega_j \left(
 -\tfrac{\omega_j}{2}
 \int 
 \vec{E}^L_i\tfrac{\Delta\varepsilon_j}{\varepsilon}\vec{E}^R_j
 \right)\equiv 2\omega_j\kappa_{ij}
\end{align} 
Assuming that the structure is approximately symmetric in the $\hat{y}$ direction (i.e., under exchanging $i\leftrightarrow j$), we have $\kappa_{ij} \approx \kappa_{ji} \approx \kappa$.
In the limit of low losses/gain, $\gamma\ll\Omega_\mathrm{EP}$, we obtain \eqref{TCMTrestriction}:
\begin{align}
\hat{A} = V^T\hat{\mathcal{A}}\,U =
\left( \begin{array}{cc}
(\omega_\mathrm{EP}-i\eta)^2 & 2\Omega_\mathrm{EP}\kappa \\
2\Omega_\mathrm{EP}\kappa & (\omega_\mathrm{EP}+i\eta)^2
 \end{array} \right).
\end{align}
where $U$ is the matrix whose columns are $\vec{E}^R_1$ and $\vec{E}^R_2$ and
$V$ is the matrix whose columns are $\vec{E}^L_1$ and $\vec{E}^L_2$


\appendixsec{Giant LDOS   enhancement in active periodic systems}{Appendix4}

\begin{figure}[b]
                 \centering\includegraphics[width=0.8\textwidth]{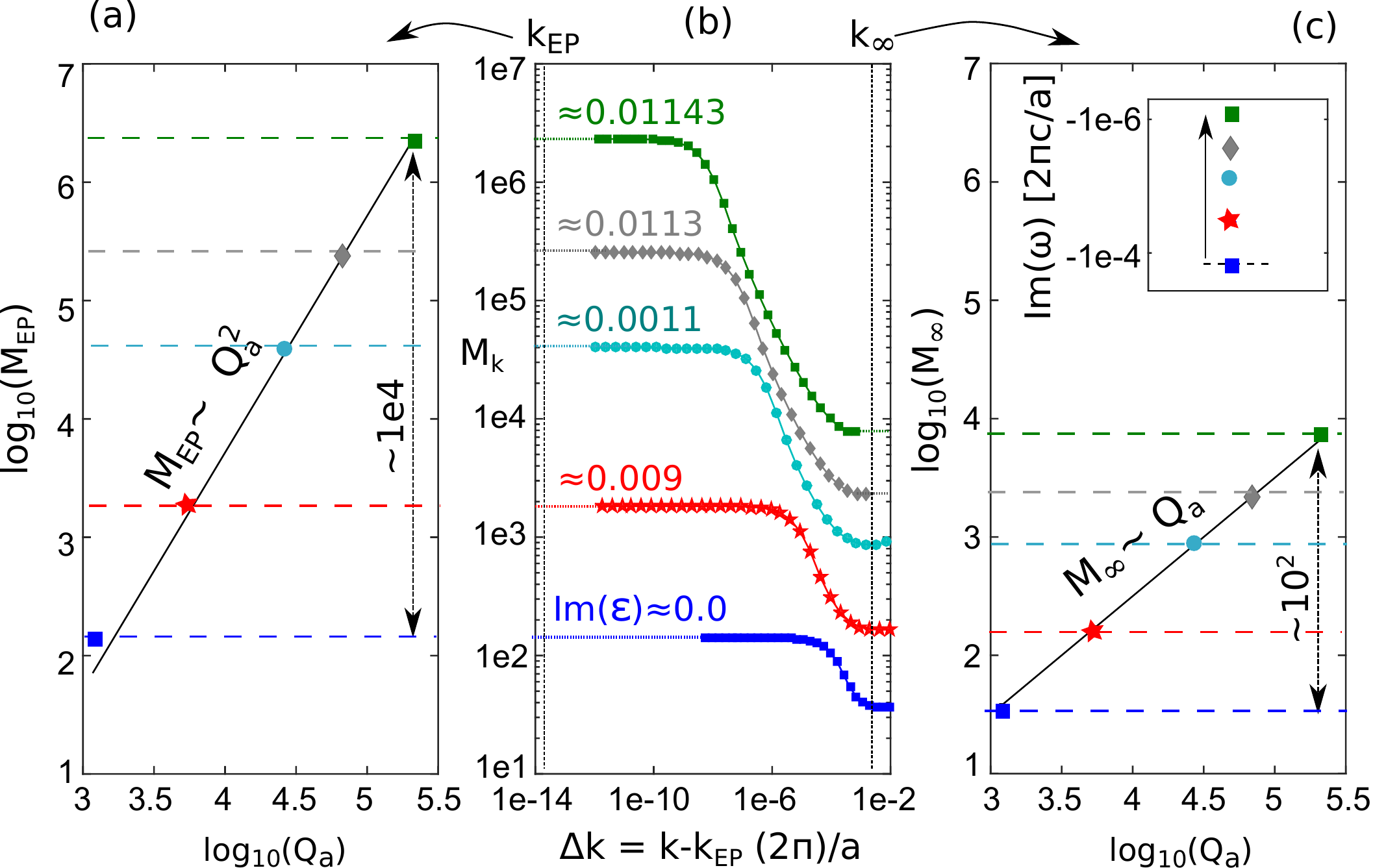}
                                   \caption{LDOS enhancement in active periodic waveguides.   
Middle:  LDOS peak ($M_k$) vs.~$\Delta k$ for increasing amounts of   gain in  the waveguide from \figref{slab-passive}(a).
Left:  $\log_{10}(M_{\mathrm{EP}})$  vs. $\log_{10}(Q_a)$ for the data from the middle panel  near $k_\mathrm{EP}$, showing quadratic scaling with $Q_a$. Right: $\log_{10}(M_\infty)$ vs. $\log_{10}(Q_a)$ for the data from the middle panel  at $k_\infty$, showing linear scaling with  $Q_a$.
Inset: Eigenvalues move vertically in the complex plane  upon adding gain.}
\label{fig:Qsqr-slab}
  \end{figure}

In this section, we demonstrate substantial  LDOS enhancements by forcing EPs in periodic waveguides with significant amounts of gain. 
Similar to \secref{Plasmonics} in the text, we  consider a  waveguide with periodic index modulation along $\hat{x}$ and outgoing boundary conditions in the transverse $y$-direction [\figref{slab-passive}(a)]. 
The  parameters of the system, prior to adding gain, are $\varepsilon_1 = 12$,  $\varepsilon_2 \approx14.43$,  and $d = 0.4807a$.

\figrefbegin{Qsqr-slab} shows our numerical results. At each wavevector, we compute  the LDOS$_k$ peak value ($M_k$) vs. deviation from the EP ($\Delta k\equiv k-k_\mathrm{EP}$) when adding  increasing  amounts of gain to the waveguide (i.e., fixing $Q_p$ while increasing $Q_a$). 
The LDOS is evaluated at $\vec{r}_0$ [\figref{slab-passive}(a)] by direct inversion of Maxwell's equations.
We show here  enhancements of $\approx400$, while higher values can easily be obtained;  the enhancement is essentially not bounded in this computational model. However, in reality, it  is bounded  by quantum noise near threshold~\cite{pick2015ab}.
 The  side panels in the figure  demonstrate  that  the LDOS peak scales quadratically with $Q_a$  near the EP [\figref{Qsqr-slab}(a)] and linearly with  $Q_a$ away from the EP [\figref{Qsqr-slab}(c)].  Finally, we note that when evaluating the LDOS  near the center of the computational cell (i.e., at $x\approx0$), the lineshape changes dramatically, and it actually has a minimum at the resonance frequency and two side peaks, whose amplitude scales as $ Q_a^2$ (not shown).

\appendixsec{Theoretical limit of LDOS  enhancement at an EP}{Appendix5}

In this section, we derive \eqref{perturbation} in the text. 
Let us first define an effective  mode amplitude $\left<\vec{E},\vec{E}\right>\equiv \int_C\!dx\,|\vec{E}|^2$, 
where $C$ denotes a finite region containing the geometry (e.g., the last scattering surface). 
In order to obtain an upper bound on the LDOS enhancement at the EP, we need to estimate the quantities
$
\left<\vec{E}_0^R,\vec{E}_0^R\right>/\left<\vec{J}_0^R,\vec{J}_0^R\right>$ and
$\left<\vec{E}_0^L,\vec{E}_0^L\right>/\left<\vec{J}_0^L,\vec{J}_0^L\right>$, which determine the relative magnitude of the two terms in the expansion formula for $G$ at the EP [\eqref{DoubleGreen}]. Let us decompose the complex-symmetric Maxwell's operator  into: $\hat{\mathcal{A}} = \hat{\mathcal{A}}' + i\hat{\mathcal{A}}''$, where $\hat{\mathcal{A}}' \equiv \tfrac{\hat{\mathcal{A}}+\hat{\mathcal{A}}^*}{2}$ and $\hat{\mathcal{A}}'' \equiv \tfrac{\hat{\mathcal{A}}-\hat{\mathcal{A}}^*}{2i}$, and the asterisk denotes complex conjugation.  
In many cases of interest, one can assume $\left\Vert \hat{\mathcal{A}}''\right\Vert\ll\left\Vert \hat{\mathcal{A}}'\right\Vert$ (under an appropriate matrix norm). In such cases,  one can  use defective perturbation theory (\subsecref{DegenPert}) to expand the eigenmodes $\vec{E}^R_\pm$ and eigenvalues $\omega_\pm$ of $\hat{\mathcal{A}}'$ in terms $\vec{E}^R_0$, $\vec{J}^R_0$ and $\lambda_\mathrm{EP}$. 
Using   \eqref{ExpandVec}, we obtain
\begin{align}
&\vec{E}_0^R \approx (\vec{E}_+^R+\vec{E}_-^R)/2 
\label{eq:pertResult1}\\
&\vec{J}_0^R \approx (\vec{E}_+^R-\vec{E}_-^R)/(2\lambda_1p^{\frac{1}{2}}).
\label{eq:pertResult2}
\end{align}
Since $\hat{\mathcal{A}}'$ is a real Hermitian operator, it has real and orthogonal eigenvectors $\vec{E}_\pm$. 
Using \eqsref{pertResult1}{pertResult2}  and assuming $\left<\vec{E}_+,\vec{E}_+\right>\approx \left<\vec{E}_-,\vec{E}_-\right>$ and $\left<\vec{E}_-,\vec{E}_+\right>=0$, one obtains
\begin{align}
&\left<\vec{E}_0^R,\vec{E}_0^R\right>\approx 
\left<\vec{E}_+,\vec{E}_+\right>/2\nonumber\\
&\left<\vec{J}_0^R,\vec{J}_0^R\right>\approx
\left<\vec{E}_+,\vec{E}_+\right>/2|\lambda_1|^2p.
\end{align}
Substituting the explicit perturbative expansion \eqref{AnsVec}, we find 
\begin{align}
\frac{\left<\vec{E}_0^R,\vec{E}_0^R\right>}{\left<\vec{J}_0^R,\vec{J}_0^R\right>} \approx|\lambda_1|^2p
\approx
\left|
\frac{(\vec{E}^L_0,\hat{\mathcal{A}}'',\vec{E}^R_0)}{(\vec{J}_0^L,\vec{E}_0^R)},
\right|
\end{align}
(with an equivalent expression  for the left eigenvector). 
Next, recall the definition of  $\hat{\mathcal{A}}\equiv\tfrac{1}{\varepsilon}\nabla\times\nabla\times$. In the limit of low losses, $\left\Vert\mathrm{Im}\,\varepsilon/\varepsilon\right\Vert
\ll1$, we can approximate~$\hat{\mathcal{A}}''\approx-\tfrac{\mathrm{Im}\,\varepsilon}{\varepsilon^2}\,\nabla\times\nabla\times\,$~and, consequently
\begin{align}
&
\Scale[0.95]{
|(\vec{E}_0^L,\hat{\mathcal{A}}'',\vec{E}_0^R)|\approx
\left|\omega_\mathrm{EP}^2 \int \vec{E}_0^L(\mathrm{Im}\,\varepsilon/\varepsilon)\vec{E}_0^R\right|= }
\nonumber\\
&
\Scale[0.95]{
\left| \omega_\mathrm{EP}^2\int\!\mathrm{Im}\,\varepsilon \,{\vec{E}_0^R}^2\right|\approx
\left|\omega_\mathrm{EP}^2\int\!\mathrm{Im}\,\varepsilon \,{|\vec{E}_0^R|}^2\right|\leq
|\omega_\mathrm{EP}|^2\,\mathrm{max}\left|\mathrm{Im}\,\varepsilon\right|,}
\label{eq:calculation}
\end{align}
where, in going from the first to the second line, we  used the relation $\vec{E}_0^R = \varepsilon^{-1}\vec{E}_0^L$ (which holds for Maxwell's eigenvalue problem) and, in the following  approximations,  we used the property $\vec{E}_0^2\approx|\vec{E}_0|^2$ (valid for  low-loss systems) and the  normalization condition $\int|\vec{E}_0|^2=1$. This completes the proof of \eqref{perturbation}.


\section*{Funding}
Army Research Office(ARO) (W911NF-13-D-0001);  U.S. Department of Energy (DOE) (DE-SC0001299);   National Science Foundation (NSF) (DMR-1307632, DMR-1454836);  United States–Israel Binational Science Foundation (BSF) (2013508). 

\section*{Acknowledgments}
AP, FH, and SGJ were partially supported by the Army Research Office through the Institute for Soldier Nanotechnologies under Contract No.~W911NF-13-D-0001. MS (reading and analysis of the manuscript) was supported by MIT S3TEC EFRC of the U.S. DOE under Grant No.~DE-SC0001299.  CWH was supported by the NSF grant No.~DMR-1307632. AWR was supported by the NSF grant No.~DMR-1454836. 
BZ was partially supported by the United States–Israel Binational Science Foundation (BSF) under award No. 2013508. 
We would like to thank Yi Yang,  Aristeidis Karalis, Wonseok Shin, Nick Rivera,  and David Liu for helpful discussions.


\end{document}